\documentclass[iop,twocolumn,twocolappendix,numberedappendix]{openjournal}
\usepackage{hyperref}
\usepackage{verbatim}
\usepackage{tabularx}
\usepackage{orcidlink}
\usepackage{amsmath}
\usepackage{xspace}
\usepackage{appendix}
\usepackage{graphicx} 
\newcommand{\fnl}[0]{f_{\rm NL}^{\rm loc}}
\newcommand{\eqreff}{Eq.~\eqref}
\newcommand{\secreff}{Section~\ref}

\newcommand{\figreff}{Figure~\ref}
\newcommand{\tabreff}{Table~\ref}
\newcommand{\regressis}{\texttt{regressis}\xspace}
\newcommand{\namaster}{\texttt{namaster}\xspace}
\newcommand{\pcl}[0]{\mathrm{PCL}\xspace}
\usepackage[capitalise]{cleveref}

\usepackage{xcolor}
\usepackage{textgreek}
\usepackage[utf8]{inputenc}
\usepackage[english]{babel}

\usepackage{hyperref}
\hypersetup{
    unicode, 
    colorlinks=true,
    linkcolor=blue,
    citecolor=blue,
    filecolor=linkcolor,
    urlcolor=blue,
}
\usepackage{color,colortbl}
\definecolor{linkcolor}{rgb}{0.0,0.3,0.5}
\usepackage{tensind}
\tensordelimiter{?}
\DeclareGraphicsExtensions{.bmp,.png,.jpg,.pdf}
\usepackage{verbatim}
\usepackage[normalem]{ulem}
\usepackage{orcidlink}
\usepackage{soul}

\begin{document}
\title{Assessing the large-scale angular clustering\\ of UNIONS Lyman Break Galaxies via cross-correlations\vspace{-1cm}}

\author{Constantin~Payerne\textsuperscript{1}}
\author{Christophe~Yèche\textsuperscript{1}}
\author{William~d'Assignies~Doumerg\textsuperscript{2}}
\author{Hendrik~Hildebrandt\textsuperscript{3}}
\author{Martin~Kilbinger\textsuperscript{4}}
\author{Calum~Murray\textsuperscript{4}}

\author{Thomas~de~Boer\textsuperscript{5}}
\author{Kenneth~C.~Chambers\textsuperscript{5}}
\author{Scott~Chapman\textsuperscript{6,7}}
\author{Michael~J.~Hudson\textsuperscript{8,9,10}}
\author{Alan~W.~McConnachie\textsuperscript{11}}
\author{the UNIONS collaboration}
\affiliation{
\textsuperscript{1}IRFU, CEA, Universit\'{e} Paris-Saclay, F-91191 Gif-sur-Yvette, France\\
\textsuperscript{2}IFAE, The Barcelona Institute of Science and Technology, Campus UAB, 08193 Bellaterra Barcelona, Spain\\
\textsuperscript{3}Ruhr University Bochum, Faculty of Physics and Astronomy, Astronomical\\
\textsuperscript{4}Université Paris-Saclay, Université Paris Cité, CEA, CNRS, AIM, 91191 Gif-sur-Yvette, France\\
\textsuperscript{5}Institute for Astronomy, University of Hawaii, 2680 Woodlawn
Drive, Honolulu HI 96822\\
\textsuperscript{6}Department of Physics and Astronomy, University of British Columbia, Vancouver, British Columbia, Canada\\
\textsuperscript{7}Department of Physics and Atmospheric Science, Dalhousie University, Halifax, Nova Scotia, Canada\\
\textsuperscript{8} Department of Physics and Astronomy, University of Waterloo, 200 University Avenue West, Waterloo, Ontario N2L 3G1, Canada\\
\textsuperscript{9} Waterloo Centre for Astrophysics, University of Waterloo, Waterloo, Ontario N2L 3G1, Canada\\
\textsuperscript{10} Perimeter Institute for Theoretical Physics, 31 Caroline St. North, Waterloo, ON N2L 2Y5, Canada\\
\textsuperscript{11}National Research Council Herzberg Astronomy and Astrophysics, 5071 West Saanich Road, Victoria, B.C., V8Z 6M7, Canada 6
}

\begin{abstract}
Lyman-break galaxies (LBGs), selected via the strong spectral break blueward of the Lyman limit, are powerful tracers of large-scale structure at redshifts $z>2$. In this work, we assess the feasibility of using LBGs selected from the Ultraviolet Near Infrared Optical Northern Survey (UNIONS) multi-band photometric catalog as cosmological probes of the high-redshift Universe using two-point statistics. We demonstrate that the spatially varying imaging systematics, driven by variations in PSF depth and seeing across the UNIONS footprint, limit robust measurements of the LBG auto-angular power spectrum on large scales, even after correcting the LBG field with linear or non-linear mitigation techniques. This study shows that clustering analyses of faint galaxy samples close to survey depth are challenging.
We therefore turn to cross-correlation measurements with external tracers, in particular the \textit{Planck} CMB lensing convergence and quasars from DESI DR1 and \textit{Quaia}, which are less sensitive to the angular imaging systematics. Using both data and mock catalogues, we demonstrate that the LBG-CMB lensing cross-power spectrum can be measured more robustly than the auto-spectrum, with an amplitude consistent with theoretical predictions. Residual systematics primarily manifest as excess variance at large angular scales, without introducing a significant bias in the recovered signal.
Taken together, these results establish UNIONS-selected LBGs as reliable tracers for cross-correlation cosmology at $z\sim 2.5$, and highlight cross-correlation techniques as a powerful and robust avenue for extracting cosmological information from photometric high-redshift galaxy samples in the presence of complex imaging systematics.
\end{abstract}

\maketitle
\section{Introduction}
Lyman-break galaxies \citep[LBGs;][]{Steidel1996LBG} are young, actively star-forming systems at $z > 1.5$, whose rest-frame ultraviolet spectra display a strong flux decrement blueward of the Lyman limit at $912\,\text{\AA}$. This spectral break arises from absorption by neutral hydrogen both within the galaxy --- in the stellar atmospheres and interstellar medium --- and along the line of sight through the intergalactic medium (IGM). Photons with $\lambda < 912\,\text{\AA}$ carry sufficient energy to ionise neutral hydrogen, producing a sharp intrinsic discontinuity in the spectrum. At longer wavelengths, $912\,\text{\AA} < \lambda < 1216\,\text{\AA}$, the spectrum is further shaped by a series of Lyman absorption lines --- including Ly$\alpha$ ($\lambda_{\alpha} = 1216\,\text{\AA}$, $2 \rightarrow 1$), Ly$\beta$ ($\lambda_{\beta} = 1026\,\text{\AA}$, $3 \rightarrow 1$), and Ly$\gamma$ ($\lambda_{\gamma} = 972\,\text{\AA}$, $4 \rightarrow 1$), converging to the series limit at $912\,\text{\AA}$ ($n \rightarrow \infty$) --- arising from resonant absorption by neutral hydrogen in both galaxies and the IGM, whose cumulative effect progressively suppresses the flux across this wavelength range.
 
This distinctive spectral signature makes LBGs powerful probes of galaxy formation and evolution at high redshift \citep{Steidel1996LBG,Steidel1999LBG,Giavalisco2004LBG,Reddy2008LBG,Hildebrandt2009lbg,Harikane2023lbg}, as well as valuable cosmological tracers. In particular, they can be used to map the large-scale structure of the high-redshift, matter-dominated Universe \citep{Foucaud2003lbgangularclustering,WilsonWhite2019dropout,RuhlmannKleider2024LBGCLAUDS,Ye2025lbgangularclustering} and to probe the IGM through Lyman-$\alpha$ forest absorption in their spectra \citep{Herrera2025lbglya}. Large and dense LBG samples at $z > 2.5$ are anticipated from current and forthcoming wide-field, multi-band photometric surveys with deep $u$-band coverage \citep{Crenshaw2025lbg,Payerne2025lbg,Payerne2025lbgforecast}, including the Ultraviolet Near Infrared Optical Northern Survey (UNIONS; \citealt{Chambers2016panstarrs,Ibata2017CFIS,Miyazaki2018HSC,Gwyn2025unions}), the Vera C. Rubin Observatory's Legacy Survey of Space and Time (LSST; \citealt{LSST2009whitepaper}), and the Chinese Space Station Telescope (CSST; \citealt{csst2016whitepaper}).
 
The high number densities of LBGs expected from these surveys will enable a broad range of precise cosmological measurements at high redshift, including projected angular clustering, cross-correlations with the cosmic microwave background (CMB) lensing potential, detection of the baryon acoustic oscillation (BAO) feature, and measurements of the three-dimensional power spectrum within next-generation spectroscopic programs such as DESI-II \citep{Schlegel2022DESI2} and future Stage-V facilities, including the Wide-field Spectroscopic Telescope (WST; \citealt{Mainieri2024wst}), the Spectroscopic Stage-5 Experiment (Spec-S5; \citealt{Besuner2025specs5}), and the MUltiplexed Survey Telescope (MUST; \citealt{Zhao2024must}).
 
As highly biased tracers spanning large cosmic volumes, LBG samples offer stringent tests of the standard cosmological model and its extensions \citep{Sailer2021highzcosmology,WilsonWhite2019dropout}. Their redshift coverage is highly complementary to that of existing large-scale structure tracers, enabling improved constraints on the growth of structure in the matter-dominated and the dark energy fraction over $2 < z < 6$, for instance through cross-correlations with CMB lensing \citep{WilsonWhite2019dropout,Miyatake2022lbgcmblensing}. Wide LBG samples are particularly well-suited to constrain models of inflation beyond the single-field paradigm favoured by \textit{Planck}, through measurements of primordial non-Gaussianity (PNG) imprinted in large-scale clustering \citep{Schmittfull2018fnl,Chaussidon2024fnl,Payerne2025lbg,Payerne2025lbgforecast,dAssignies2023forecast}. On smaller scales, dense LBG samples further provide sensitivity to the sum of neutrino masses \citep{Yu2018neutrinomass,dAssignies2023forecast}.
 
In this context, the Ultraviolet Near Infrared Optical Northern Sur-
vey (UNIONS, \citealt{Gwyn2025unions}) combines multi-band imaging from several telescopes in Hawaii: the Canada-France Imaging Survey (CFIS) provides $u$- and $r$-band
data from CFHT, Pan-STARRS contributes $i$- and $z$-
band imaging, and Subaru adds $z$-band coverage through WISHES and g-band coverage through WHIGS. The first cosmic shear cosmology constraints with UNIONS have recently been presented in a series of articles \citep{HervasPeters2026unionslensingcatalogue,Daley2026UNIONSBmode,Goh2026UNIONScosmologyconfigurationspace,Guerrini2026UNIONScosmologyharmonic}.  As the widest and deepest $u$-band survey currently available --- before the release of Rubin LSST Year 2 data --- UNIONS constitutes a unique open window for high-redshift LBG cosmology in the era of Stage-IV photometric surveys, through $u$-dropout LBG selection.
 
The primary goal of this study is to explore the feasibility of selecting LBGs from the 3,500 square degrees UNIONS multi-band dataset, extending the analyses of \citet{Payerne2025lbg}, which were based on artificially degraded photometry, and the forecasting framework established in \citet{Payerne2025lbgforecast}. This work represents a first step toward constructing wide-area LBG samples suitable for cosmological analysis, to characterise their uniformity across different survey strategies, as well as their redshift distributions and number densities.
 
The second objective is to evaluate the feasibility of measuring the large-scale clustering of UNIONS-selected LBGs. Probing large scales with high-redshift tracers is of particular importance for constraining local primordial non-Gaussianity, which induces a characteristic scale-dependent enhancement in the linear bias of cosmological tracers \citep{Dalal2008,Slosar2008}. This signature has been extensively exploited to constrain $f_{\rm NL}^{\rm loc}$, both through the three-dimensional power spectrum \citep{Rezaie2023,Cagliari2023,Chaussidon2024fnl} and via angular power spectra and cross-correlations with CMB lensing \citep{Krolewski2024fnlcmblensing,Fabbian2025lbgcmblensing,Climent2025fNLcross,Chiarenza2025cmblensing}. The tightest current constraint, $f_{\rm NL}^{\rm loc} = -3.6^{+9.0}_{-9.1}$, was obtained by \citet{Chaussidon2024fnl} from large-scale power spectrum measurements of DESI quasars (QSOs, $0.8 < z < 3.1$) and luminous red galaxies (LRGs, $0.6 < z < 1.1$).
 
High-redshift LBGs are expected to yield independent and competitive constraints on $f_{\rm NL}^{\rm loc}$, owing to their higher number densities relative to DESI quasars at comparable redshifts \citep{Payerne2025lbg,Crenshaw2025lbg,Payerne2025lbgforecast} and their redshift distribution spanning $z \sim 2.5$--$3.5$ \citep{2023MNRAS.521.3648D}. However, spatially varying coverages of photometric surveys -- impacting the homogeneity of survey depth and seeing -- as well as Galactic extinction affect the homogeneity of the source selection, and render clustering measurements challenging, whether relying on the projected (see e.g. \citealt{Krolewski2024fnlcmblensing}) or three-dimensional tracer's power spectrum \citep{Chaussidon2024fnl}. In the absence of spectroscopic follow-up, as is the case here, it is often more robust to probe the large-scale matter distribution through cross-correlations with CMB lensing, where uncorrelated systematics between the photometric tracer and the CMB observables are largely suppressed --- a property that is particularly valuable for studies of local PNG \citep{Krolewski2024fnlcmblensing,Chiarenza2025cmblensing}.
 
In this paper, we assess the feasibility of measuring the cross-correlation signal between UNIONS-selected LBGs and CMB lensing maps, and use the amplitude of this signal as a validation of the LBG sample. This analysis constitutes a first evaluation of the potential of UNIONS LBGs for future constraints on $f_{\rm NL}^{\rm loc}$.
 
The paper is organized as follows. In \secreff{sec:formalism_AngPow}, we present the formalism adopted to measure the clustering properties of LBGs, covering both the estimator and modeling aspects. \secreff{sec:UNIONS_data} describes the UNIONS data, the survey characteristics, and the LBG selection strategies explored in this study. In \secreff{sec:imag_syst}, we characterize the level of imaging contamination in the LBG density maps and assess several mitigation techniques, highlighting in particular that the LBG auto-spectrum is largely unsuitable for cosmological inference, whereas cross-correlations with external datasets offer a robust alternative. Finally, in \secreff{sec:cross-correlating_lbg_ext}, we present the cross-correlation signal between UNIONS LBGs and external datasets.
\section{Angular Power Spectrum: Formalism}
\label{sec:formalism_AngPow}

Throughout this paper, we measure the angular power spectrum of two projected fields, $X$ and $Y$, where each field represents either the LBG overdensity or an external overdensity field; the case $X = Y$ corresponds to the auto-angular power spectrum of field $X$. The pseudo-angular power spectrum between two cosmological fields $X$ and $Y$ is defined as
\begin{equation}
    \widehat{\pcl}_\ell^{XY} = \frac{1}{2\ell + 1} \sum_{m=-\ell}^{\ell} \left| \widehat{a}_{\ell m}^{X,*} \widehat{a}_{\ell m}^Y\right|^2,
    \label{eq:pcl_ell_est}
\end{equation}
where the spherical harmonic coefficients are given by
\begin{equation}
    \widehat{a}_{\ell m}^{X} = \int d^2\vec{\theta}\ m(\vec{\theta})\,\widehat{s}^{X}(\vec{\theta})\,Y_{\ell m}(\vec{\theta}),
\end{equation}
with $\widehat{s}^{X}(\vec{\theta})$ denoting the field $X$ --- either a galaxy overdensity field or the CMB lensing convergence $\kappa(\vec{\theta})$ --- and $m(\vec{\theta})$ the survey mask over which $X$ and $Y$ are defined. The binned decoupled angular power spectrum $\widehat{C}_b^{XY}$ is then obtained via
\begin{equation}
    \widehat{C}_b^{XY} = \mathcal{M}_{bb'}^{-1}\, B_{b\ell}\,\widehat{\pcl}_{\ell}^{XY}
     \label{eq:cl_b}
\end{equation}
where $B_{b\ell}$ is the $\ell$-to-$b$ binning matrix, and $\mathcal{M}_{bb'} = B_{b\ell}\,\mathcal{M}_{\ell \ell'}\,B^{-1}_{\ell'b'}$, and $\mathcal{M}_{\ell \ell'}$ is the mode-coupling matrix encoding the effects of partial sky coverage,
\begin{equation}
\label{eq:modemixing_matrix}
    \mathcal{M}_{\ell \ell'} = \frac{2\ell'+1}{4\pi} \sum_{\ell''} (2\ell''+1)\, \pcl^{\rm mask}_{\ell''}
\begin{pmatrix}
\ell & \ell' & \ell'' \\
0 & 0 & 0
\end{pmatrix}^2,
\end{equation}
with $\pcl^{\rm mask}_{\ell''}$ the pseudo-angular power spectrum of the survey mask $m(\vec{\theta})$. Throughout this work, we make use of the \namaster code\footnote{\url{https://github.com/LSSTDESC/NaMaster}}~\citep{Alonso2019namaster,Cornish2026namaster}, which provides a unified framework for computing angular cross-power spectra, the mode-coupling matrix $\mathcal{M}_{\ell \ell'}$, and the mitigation of imaging systematics.
 
Equation~\eqref{eq:cl_b} provides an estimator for the binned decoupled angular power spectrum between two projected fields $X$ and $Y$, which can be related to the underlying matter power spectrum. Each field $X$ can be expressed as a weighted projection of the matter overdensity along the line of sight,
\begin{equation}
    \widehat{\delta}^X(\vec{\theta}) = \int d\chi\, q_X(\chi)\, \widehat{\delta}_m(\vec{\theta}, \chi),
\end{equation}
where $\chi$ is the comoving distance and $q_X(\chi)$ is the corresponding radial projection kernel. The ensemble-average angular power spectrum $\langle \widehat{C}_\ell^{XY}\rangle = C_\ell^{XY}$ is then given by
\begin{equation}
C_\ell^{XY} = \frac{2}{\pi} \int dk\, k^2\, P_{m}(k)\,
\Delta_\ell^X(k)\, \Delta_\ell^Y(k) + N^{XY}_\ell,
\end{equation}
where
\begin{equation}
\Delta_\ell^X(k) = \int_0^{\chi_{\rm H}} d\chi\, q_X(\chi)\, j_\ell(k\chi),
\end{equation}
and $P_m(k)$ is the matter power spectrum. The noise term $N^{XY}_b = \delta_K^{XY} N_b^X$ is non-zero only for the auto-spectrum of a given field. For a discrete point-like tracer, $N_b^X = 1/\bar{n}_X$ is the shot-noise contribution arising from the finite sampling of the underlying matter density field. For the CMB lensing map, $N_b^X = N_b^\kappa$ denotes the reconstruction noise of the lensing estimator.
 
For galaxy or quasar tracers, the projection kernel $q_X(\chi)$ is proportional to the redshift distribution of the sample, weighted by the tracer bias. For CMB lensing, $q_X(\chi)$ takes the form of the lensing convergence kernel, integrated from $z = 0$ to the last-scattering surface at $z \approx 1100$ (see Appendix~\ref{sec:modeling_ang_pow_spec}).

\section{UNIONS data}
\label{sec:UNIONS_data}
\subsection{The UNIONS Imaging Survey}
\begin{figure*}
    \centering
\includegraphics[width=0.49\linewidth]{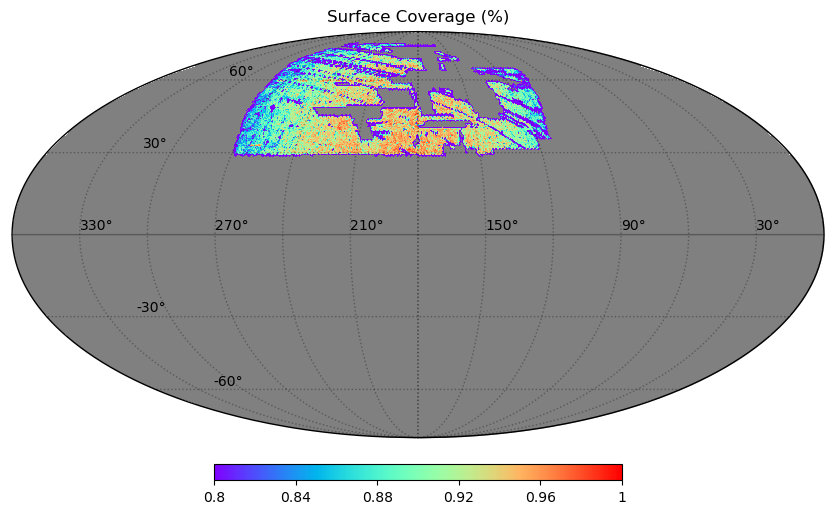}
\includegraphics[width=0.49\linewidth]{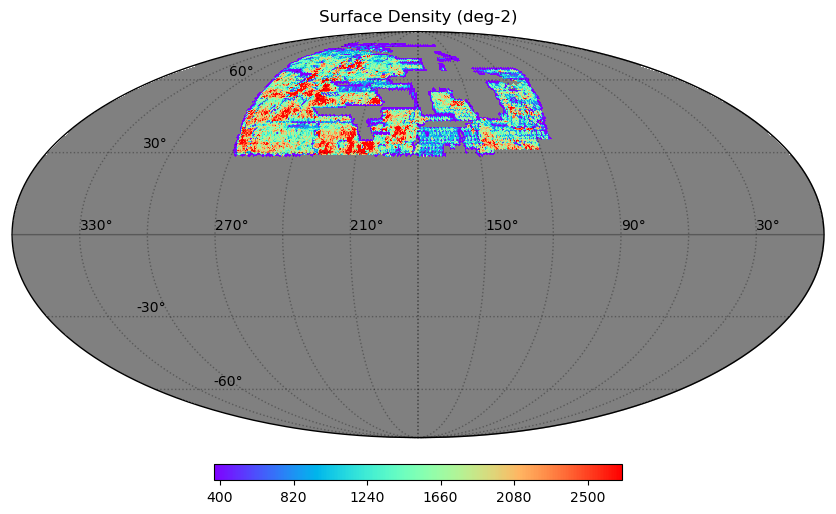}
    \caption{Left: HEALPix pixel coverage for the GAaP footprint. Right: Local density of selected $u$-dropout galaxies (in deg$^{-2}$), corrected from surface coverage.}
    \label{fig:surface_coverage}
\end{figure*}
UNIONS Multi-band object catalogs were generated using the GAaP (Gaussian Aperture and PSF; \citealt{kuijken2008,kuijken2015,kuijken2019}) algorithm, which computes photometry in the $ugri$ (or $ugriz$) bands based on the detection location in the CFIS $r$ band. Since our LBG selection relies on $ugr$ magnitudes, we use the UNIONS $ugri$-GAaP catalog, which is substantially larger than the $ugriz$-GAaP catalog. The coverage of the UNIONS $ugri$-GAaP tiling is roughly 3,580 square degrees.

A series of masks is applied to account for bad pixels, star spikes, and satellite trails (based on $r$-band images). For each pixel of the HEALPix \citep{healpix2004} map, we estimate its fractional coverage by generating random points across the sky at a high density of 15,000 points per square degree and comparing the local densities before and after masking. The surface coverage is illustrated in the left panel of \figreff{fig:surface_coverage}. Within the footprint defined by the combination of the different UNIONS sub-surveys, line-shaped features are visible due to some GAaP tiles being coherently missed; these will be corrected in the next version of the GAaP catalog. The HEALPix coverage is minimal at the edges of the footprint, as pixels surrounding the survey area may be only partially covered by UNIONS GAaP tiles (a purely pixelization effect). Overall, the HEALPix coverage gradually decreases from the center of the UNIONS GAaP footprint toward the edges, reflecting proximity to the Galactic plane. The per-pixel fractional coverage $f_i$ of each HEALPix pixel is used to locally correct the counts of tracers, $N_{\rm obs}^i$, by rescaling them as $N_{\rm corr}^i = N_{\rm obs}^i / f_i$. After accounting for masking, the \textit{effective} UNIONS GAaP area is 2,800 square degrees.

\subsection{Dropout selection of high-redshift galaxies using UNIONS GAaP magnitudes}
We consider $23.5 < r < 24.2$ galaxies, that we select with a $u$-dropout criteria \citep{RuhlmannKleider2024LBGCLAUDS}
\begin{align}
    (i)&\ u - g > 0, \\
    (ii)&\ -0.5 < g - r < 1.2, \\ 
    (iii)&\ [u - g > 2.2 \times (g - r) + 0.32] \ \cup \\
    &\ [u - g > 0.9 \ \cap \ u - g > 1.6 \times (g - r) + 0.75].
\end{align}
where (i) represents the $u$-dropout selection, which requires the flux in the $g$ band to exceed that in the $u$ band. Moreover, (ii) and (iii) are necessary to avoid contamination from the stellar locus.
We apply additional quality cuts on the magnitudes of interest: $\mathrm{err}(u,g,r) > 0$, $\mathrm{err}(u) < 3$, $\mathrm{err}(g) < 3$, $\mathrm{err}(r) < 1$, and $|u-g| < 3$, $|g-r| < 3$, ensuring that galaxy colors lie within a reasonable range around zero. 

\begin{figure*}
       \centering
\includegraphics[width=1\linewidth]{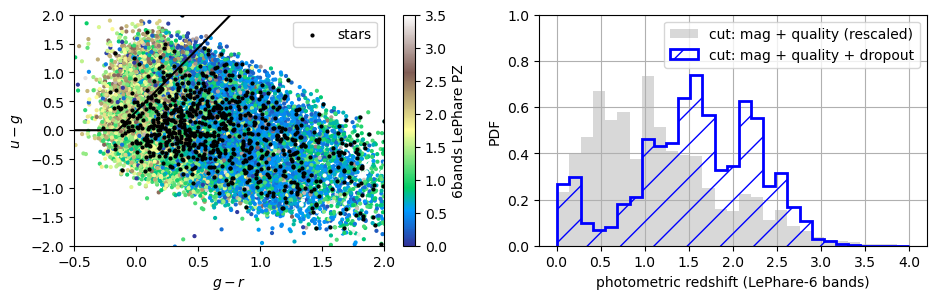}
       \caption{Left: color-color diagram of UNIONS galaxies, with $u$-dropout selected objects shown in blue. Red points correspond to stars (specified with the flag \texttt{OBJ\_TYPE\_hsc==1} in the CLAUDS+HSC catalogs).  
Right: Photometric redshift distribution of the $u$-dropout galaxies, compared to the redshift distribution of all galaxies before applying the $u$-dropout selection.}
    \label{fig:diagnostic_nz_xmm_udropout}
\end{figure*}

We leverage the UNIONS coverage of the XMM-LSS field (approximately 2 square degrees), where $ u$-dropout-selected galaxies are matched with CLAUDS+HSC-detected objects \citep{Desprez2023CLAUDSHSCSSP}. The latter benefit from precise photometric redshifts derived from six optical bands using the LePhare template-fitting code \citep{Arnouts1999,Arnouts2002,Ilbert2006}. In the left panel of \figreff{fig:diagnostic_nz_xmm_udropout}, we show the color-color diagram ($u-g$ versus $g-r$) of UNIONS GAaP objects within the $r$-band range $23.5 < r < 24.2$, after applying the quality cuts, along with the $u$-dropout selected objects.
The right panel of \figreff{fig:diagnostic_nz_xmm_udropout} displays the resulting LePhare photometric redshift distribution of these objects. Due to overlap between low-redshift ($z<1.5$) and higher-redshift ($z>2$) objects in the color-color diagram, the $u$-dropout selection exhibits a relatively broad redshift kernel, peaking at $z \approx 1.5$–$2$, already mentioned in \citet{Payerne2025lbgforecast}. The excess probability at redshifts below $z=0.5$ corresponds to Emission Line Galaxies, selected through $u$-dropout caused by the Balmer Break at 3646 $\AA$.  The shape of the redshift kernel of a similar $u$-dropout LBG selection over the full UNIONS GAaP-$ugri$ footprint was validated in a previous analysis \citep{Payerne2025lbgforecast} using clustering-redshift methods \citep{Menard2013Cz,dAssignies2025Cz}, leveraging DESI DR1 data over the full redshift range $0 < z < 3.5$ and covering approximately 1,000 square degrees common to UNIONS and DESI DR1.

We apply this selection to the full GAaP footprint, and in the right panel of \figreff{fig:surface_coverage}, we show the local density of the $u$-dropout sample (per square degree), after correcting for surface coverage, giving an average LBG surface density of $n_{\rm gal}=$1400 per square degree. We see that, even after correcting the density from the impact of the survey coverage, large-scale fluctuations imprint the LBG distribution, revealing the presence of non-astrophysical imaging systematics that we address in the next section.

\section{Angular Systematics in the LBG sample}
\label{sec:imag_syst}
The LBG angular distribution is locally affected by survey systematics, as spatial variations seem to be correlated with specific survey designs. In this section, we investigate how spatially varying imaging features--either from the UNIONS survey strategy or from astrophysical effects -- affect the observed cosmological angular distribution of Lyman-Break Galaxies across the UNIONS footprint. As a recap of the section, we find that the large-scale modes of the LBG auto-spectrum remain largely unusable for cosmological analyses, as imaging systematics are difficult to fully remove. This is a limitation common to most studies using cosmological samples selected on wide-field photometric surveys. The dependency of the tracer local density on the survey design features (depth, seeing) should not be interpreted as a deficiency of the underlying survey data or photometric measurements (see e.g., \citealt{Krolewski2023}).

\subsection{Dependence of the LBG density on the imaging features}

The different UNIONS sub-surveys (CFIS, Pan-STARRS, WISHES and WHIGS) are not perfectly homogeneous in depth and image quality, reflecting variations in observing strategy and observing conditions across the surveys. The PSF depth and seeing coverage for the $u$, $g$, $r$ bands are shown in Appendix \ref{app:unons_templates}, in \figreff{fig:unions_depth} and \figreff{fig:unions_seeing}, respectively. PSF depth fluctuations can impact LBG selection when applying color and magnitude cuts. Similarly, PSF size (or seeing) exhibits spatial inhomogeneities across the UNIONS footprint, affecting the quality of derived detection catalogs and, ultimately, clustering measurements. Even after masking stars and correcting GAaP magnitudes for $E(B-V)$ extinction, the presence of stars and extinction can still degrade image quality—for example, by locally increasing background photon noise—which may influence the detection of objects.

\begin{figure*}
    \centering
    \includegraphics[width=0.9\linewidth]{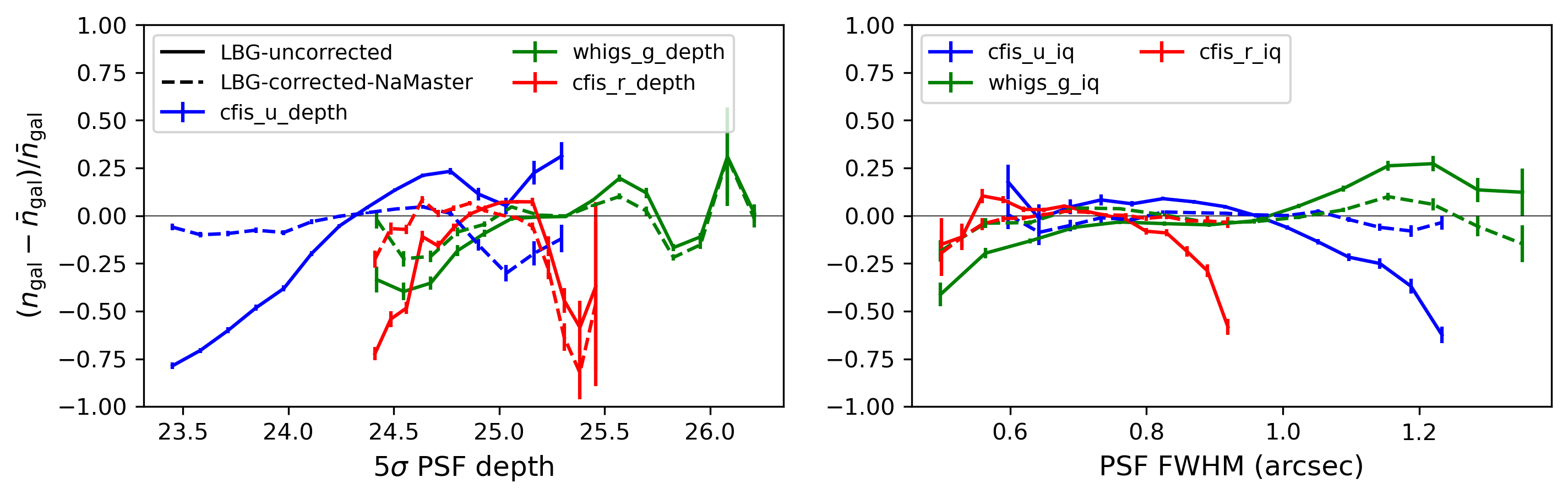}
    \includegraphics[width=0.9\linewidth]{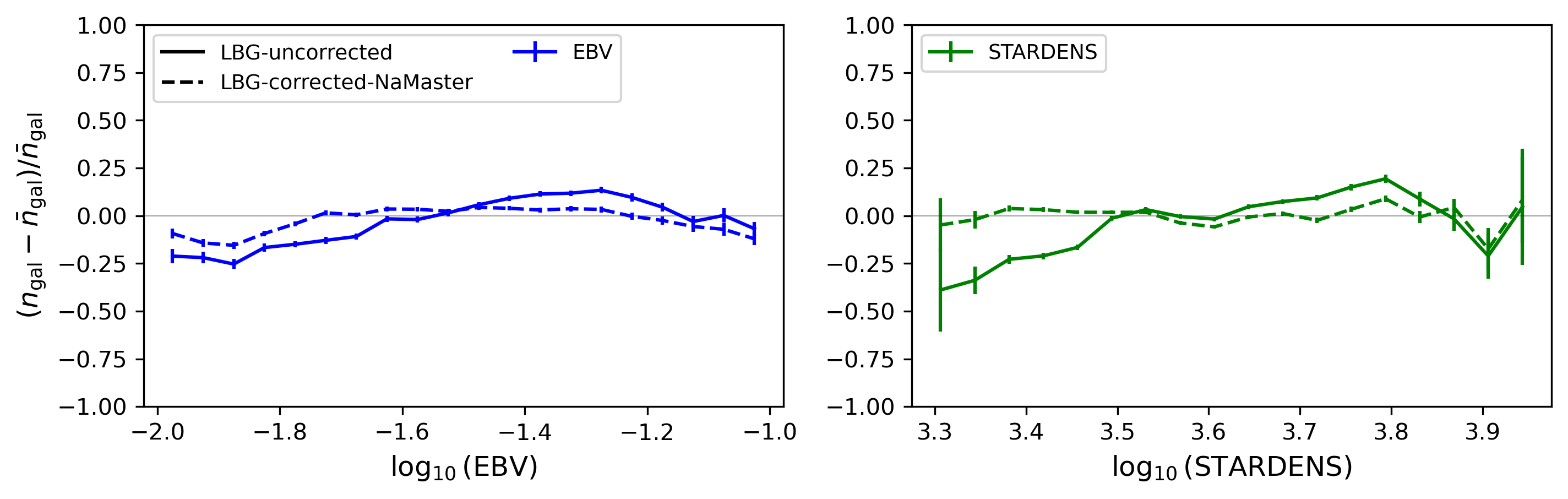}
    \caption{Upper panels: LBG overdensity as a function of UNIONS survey characteristics (left: $5\sigma$ PSF depth, right: PSF size in arcsec).  
Lower panels: Overdensity as a function of external templates (left: Galactic extinction $E(B-V)$, right: stellar density).  
Solid lines correspond to the uncorrected LBG field, while dashed lines show the results after deprojection with \namaster.}
    \label{fig:overdensity_VS_template_namaster}
\end{figure*}

The two upper panels of \figreff{fig:overdensity_VS_template_namaster} show the LBG $u$-dropout overdensity as a function of UNIONS survey characteristics: $5\sigma$ PSF depth (left) and seeing (right). The two lower panels show the overdensity as a function of $E(B-V)$ \citep{Schlegel1998ebv} and stellar density \citep{Gaia2023stardens}. It is evident that the LBG density is significantly affected by the UNIONS survey strategy; for example, the local LBG density is notably lower in pixels with shallow $u$-band coverage or large $u$-band PSF size.
We also note a mild positive correlation between the local LBG density and the external templates.
For several pixels, however, it is challenging to statistically correct for these effects, as the required rescaling can be substantial—the observed LBG counts sometimes need to be increased by factors of $5$–$10$. To mitigate this, we exclude pixels with extreme values of depth or seeing. The "cleaned" footprint, in terms of UNIONS PSF depth and size, is defined as $23 < u_{10\sigma} < 24.5$, 
$24 < g_{10\sigma} < 25.1$, 
$23.5 < r_{10\sigma} < 24.5$, 
$0.5 < u_{\rm PSFsize} < 1.2$, 
$0.5 < g_{\rm PSFsize} < 1.2$, 
$0.5 < r_{\rm PSFsize} < 1$. We also remove by hand some "bad" photometric regions (in terms of PSF and seeing) appearing in the HEALPix maps in Appendix \ref{app:unons_templates}; we remove (i) the Dec  $>75$ deg part of the footprint (poor $u$-band seeing) (ii) the Ra $\in [150,180]$ and Dec $\in [20,45]$ region, which shows specific repetitive patterns of image quality in the $g$-band, affecting "by eye" the LBG density (see the left panel of \figreff{fig:surface_coverage}). Such cleaning introduces additional missing regions in the footprint for an effective area of 2,400 square degrees.

To mitigate the effect of angular imaging systematics in the observed LBG density, we adopt the linear regression method implemented in \namaster \citep{Alonso2019namaster,Cornish2026namaster}, which has been used in previous clustering analyses \citep{Alonso2023quaia,Fabbian2025lbgcmblensing}.  This method models the observed overdensity field, $\delta^{\rm obs}_{\rm LBG}(\boldsymbol{\theta})$, as a linear combination of the true underlying density field and a set of systematic templates, $t^i(\boldsymbol{\theta})$. The corrected overdensity field is then expressed as
\begin{equation}
    \delta_{\rm LBG}^{\rm corr}(\boldsymbol{\theta}) = \delta^{\rm obs}_{\rm LBG}(\boldsymbol{\theta}) - t^i(\boldsymbol{\theta})\,F_{ij}\,\int d\boldsymbol{\theta}'\, t^j(\boldsymbol{\theta}')\,\delta^{\rm obs}_{\rm LBG}(\boldsymbol{\theta}') \,,
\end{equation}
The matrix $F_{ij}$ accounts for correlations between different systematics and is defined as the inverse of the template auto-correlation matrix,
\begin{equation}
    (F^{-1})_{ij} = \int d\boldsymbol{\theta}\; t^i(\boldsymbol{\theta})\, t^j(\boldsymbol{\theta}) \,.
\end{equation}
This method assumes that contamination from imaging systematics is linear in the templates and that the selected templates adequately capture the dominant observational effects. The \namaster de-projection procedure removes some power on large scales—a phenomenon referred to as angular mode removal—and also modifies the tracer white noise (called de-projection noise bias). After de-projection, the corrected overdensity is shown as a function of the different imaging templates in \figreff{fig:overdensity_VS_template_namaster}. The procedure significantly reduces correlations between the LBG overdensity and the imaging templates, although the correction is less effective for PSF depth features, which exhibit strong non-linearities. We emphasize that \namaster dep-projection is not the unique method for correcting the tracer's spatial variations from inhomogeneities in instrumental and astrophysical quantities. We adopt this rather conservative approach, being aware that other methods are known to capture more complex, non-linear dependencies between the tracer density spatial variations and survey templates, by e.g., fitting the survey selection function form $k$-means clustering approach and weighting mocks for unbiased clustering measurements (this method was explicitly developed for correcting LBGs overdensity from PSF depth, seeing and galactic extinction by \citealt{Morrison2015mitigation}). Or else using a Random Forest to learn the pixel weighting scheme that reproduces the observed tracer-template dependency (such as with the \regressis code, see \citealt{Chaussidon2022qsoang}). In the next sections, we will compare the performance of the \namaster linear de-projection technique with the \regressis code.
\subsection{LBG angular power spectrum: Impact of the \namaster deprojection}

\begin{figure*}
    \centering
    \includegraphics[width=1\linewidth]{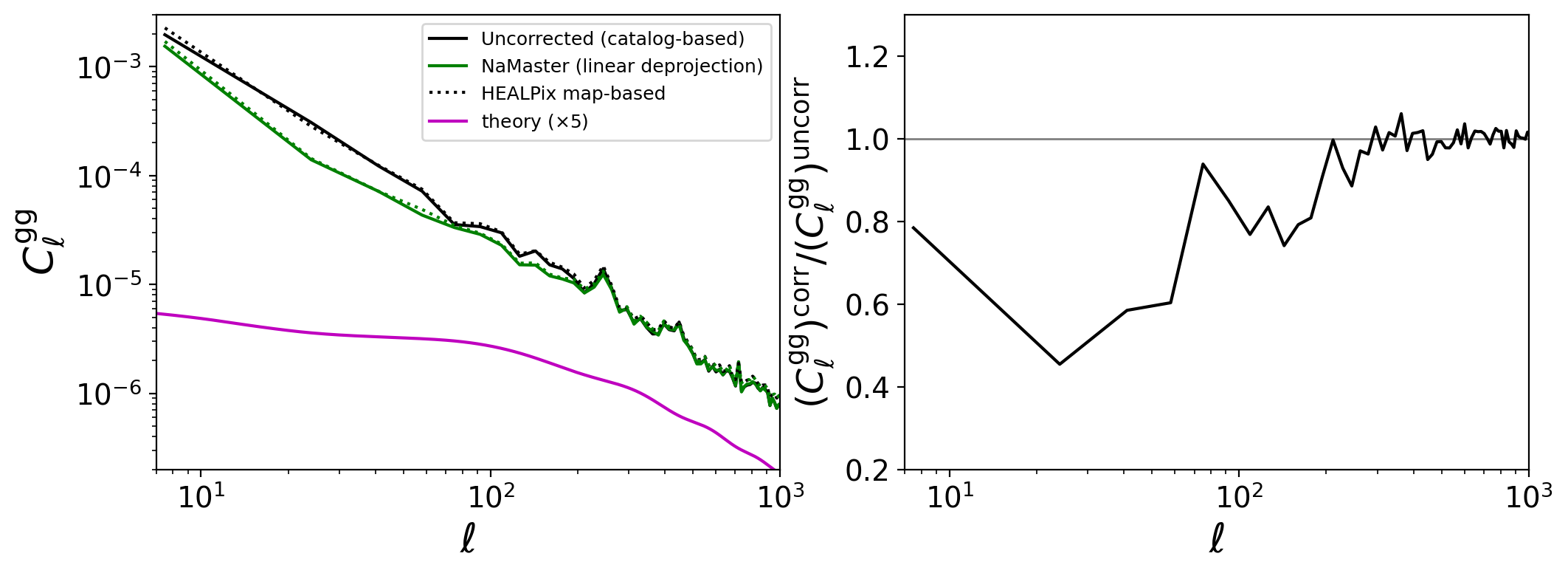}
    \caption{\textbf{Left:} Angular power spectrum of the LBG density field before and after mitigation with \namaster (note that neither angular mode removal nor deprojection noise bias has been applied). Dots refer to the catalog-based approach in \namaster, when full lines denote for the HEALPix-based \namaster method. The curve at the bottom of the plot corresponds to a fiducial model for the LBG angular clustering amplitude, mutliply by 5 so it fits in the frame of the figure. 
\textbf{Right:} Comparison of the corrected LBG auto-spectrum to the uncorrected spectrum.}
    \label{fig:Angpow_Cell_gg_lbgudropout}
\end{figure*}

In the left panel of \figreff{fig:Angpow_Cell_gg_lbgudropout}, we show the angular power spectrum of the $u$-dropout LBG sample before and after applying the \namaster de-projection. As expected, the correction reduces power at large scales relative to the uncorrected spectrum. The right panel shows the ratio of corrected to uncorrected power, highlighting a substantial reduction of roughly 50$\%$ at $\ell \lesssim 100$, with negligible impact at $\ell \gtrsim 300$. The scales of primary interest in this work are $\ell < 300$.
We overplot a fiducial theoretical prediction of the expected clustering amplitude, with ingredients (bias model, magnification bias) detailed in \citet{Payerne2025lbg} for the same LBG selection. Although this model may not be exact to describe the clustering amplitude of the selected LBGs and should not be taken as a reference for quantifying the impact of angular imaging systematics (and their mitigation), it helps demonstrate qualitatively that the measured power clearly exceeds the theoretical expectation by several orders of magnitude, indicating residual contamination from imaging systematics. Even after mitigation, a qualitative agreement between the corrected spectrum and the fiducial model is not recovered at all, consistent with previous studies on the angular clustering of photometric samples: the largest angular scales in photometric samples are challenging to measure robustly, particularly when inhomogeneities in survey strategy—arising from different telescopes, varying depth, and seeing—affect tracer selection. Furthermore, the LBG sample used in this analysis is a faint, magnitude-limited sample whose limiting magnitude approaches the survey depth in the $r$-band. This places the sample in a regime where the observed galaxy number density is highly sensitive to subtle spatial variations in survey depth and image quality, as small changes in the local photometric conditions directly affect the completeness of the source selection and hence the detected number counts. We emphasize that the imaging systematics affecting the feasibility of measuring the LBG auto-spectrum are not indicative of a deficiency in the UNIONS photometric pipeline. the challenge arises instead from the nature of the selected sample itself, such as a faint, magnitude-limited sample whose flux limit approaches the survey depth means that even subtle, spatially coherent variations in observing conditions --- such as small fluctuations in PSF depth or seeing --- translate into significant fluctuations in the detected number counts.
While these systematics significantly impact autocorrelation measurements, they are much less severe for cross-correlations with external datasets that do not share the same systematics, such as CMB lensing or QSO samples. Accordingly, this paper focuses on cross-correlations, which provide a more reliable probe of the large-scale distribution of LBGs. We still note that, even if cross-correlation signals are less sensitive (this is the purpose of the next section). The LBG auto-spectrum (before or after correction) still shows a high clustering amplitude, that will impact, in the end, the variance of the cross-spectrum estimators. This will be discussed in the next section too.

\subsection{The power of cross-correlation: Investigation of the impact of imaging systematics with mocks}
Although the LBG auto-spectrum is unsuitable for cosmological analyses due to residual contamination that would require a better purity treatment beyond the scope of this work, cross-correlating UNIONS LBGs with external datasets helps mitigate the impact of imaging systematics, since these systematics are largely uncorrelated between independent probes. The effect of imaging systematics, as well as the performance of the \namaster de-projection method, can be tested using simulated datasets. The procedure to generate a contaminated LBG map that is correlated with an external projected map is as follows:
\begin{enumerate}
\item \textbf{Compute contamination weights:} Starting from an originally contaminated LBG map, we employ the independent code \regressis \citep{Chaussidon2022qsoang}, which has also been used to mitigate angular imaging systematics for DESI tracers \citep{Chiarenza2025cmblensing,Chaussidon2024fnl,Krolewski2024fnlcmblensing}. The output of \regressis is a HEALPix map of systematic weights, $w_{\rm sys}$, to be applied to the contaminated pixels. The local tracer density is modeled as $\widehat{N}_i^{\rm obs} = N_0 \, (1 + \widehat{\delta}_i) \, F(s_i)$
where $N_i^{\rm obs}$ is the observed galaxy count in HEALPix pixel $i$, $f_i$ is the pixel surface coverage, $N_0$ is estimated as $
\widehat{N}_0 = \langle f_i^{-1} \widehat{N}_i^{\rm obs} \rangle$
and $F(s_i)$ represents the dependence of the tracer density on observational systematics. The function $F(s_i)$ is inferred using either a Linear regression or a Random Forest (RF) approach, employing $k$-fold cross-validation over the available pixels. The resulting systematic weight $w_{\rm sys}$ is defined such that $\widehat{N}_i^{\rm corr} = w_{\rm sys} \, \widehat{N}_i^{\rm obs}$. In this work, we adopt the Random Forest regression to capture complex, non-linear dependencies of the tracer angular distribution on imaging systematics. 
\\
\item \textbf{Generate uncontaminated correlated fields:} We generate two correlated Gaussian fields with the \texttt{healpy} package, the first one corresponding to the LBG overdensity map and the other corresponding to another projected map with its own redshift kernel (for the following, it will be associated to a mock CMB lensing map, with an associated noise spectrum from \textit{Planck} PR4 \citealt{Carron2022PR4Plancklensing}). One issue is that we lack a clear understanding of the true LBG redshift distribution, bias, and magnification. We adopt a fiducial model that is taken from the previous UNIONS LBG forecast study \citep{Payerne2025lbgforecast}, when the LBG population peaks at a redshift of $2.8$, with a bias of $3.5$, slightly different from the UNIONS LBG characteristics shown in the previous section (where the LBG redshift distribution was measured in XMM-LSS), but does not change the consistency of the procedure in practice. The LBG overdensity map is Poisson-sampled at a very high density of 15,000 deg$^{-2}$ to down-sample it to the desired contamination level (see below). \\
\item \textbf{Contaminate the LBG mock map:}  The next step consists of downsampling the high-density LBG mock map to account for partial sky coverage and the inverse Random Forest \regressis weights. The map is reduced to a density of $1{,}500~\mathrm{deg}^{-2}$, roughly ten times lower than the initial value, to match a typical observed LBG density. We note that no contamination is applied to the external map, in this case, the CMB lensing field. 
\end{enumerate}
As a recap, starting from an originally contaminated map, the procedure outlined above allows us to (Step 1:) extract the systematic weight map $w_{\rm sys}$ from a given density map, (Step 2:) generate clean LBG mocks correlated with a CMB lensing field, and (Step 3:) contaminate clean mock LBG density maps. The performance of the contamination–decontamination procedure can then be tested by applying \namaster to the contaminated mocks, allowing us to assess whether the input “true” signal is correctly recovered.
\subsubsection{Validation of the method: An analytical contamination model}
We first consider an analytical contamination model based on \citet{Awan2025lsst}. We generate an initial mock masked Gaussian random field $\widehat{\delta}_{\mathrm{LSS}, i}$ over the UNIONS GAaP footprint, with an input angular power spectrum $C_\ell^{gg}$, we contaminate the overdensity field through
\begin{equation}
     (1 + \widehat{\delta}_{\mathrm{obs}, i})=(1 + \widehat{\delta}_{\mathrm{LSS}, i})\prod_{t=1}^{n_t} (1 + A_t[\widehat{\delta}_{\mathrm{t},i}]^\alpha) 
     \label{eq:linear_contamination_model}
\end{equation}
where $\widehat{\delta}_{\mathrm{t},i}$ is the template overdensity in pixel $t$, where $n_t=8$ considered templates, and $A_t$ are correlation amplitudes\footnote{to be Gaussian random variables to mimic, on average, the overall correlation amplitude measured between the data and the templates}. For $\alpha=1$ in \eqreff{eq:linear_contamination_model} (corresponding to the model used in \citealt{Awan2025lsst}), the contamination is linear, while for $\alpha=3$ in \eqreff{eq:linear_contamination_model}, it introduces non-linear effects. This model, although being qualitative, enables us to assess the performance of the mitigation methods since we know the contamination recipe. The field is then Poisson-sampled to include shot noise in the measurements.  We then extract an RF \regressis weight map from the contaminated mock (as detailed in Step 1), which is applied to clean mocks generated with the same $C_\ell^{gg}$ (Step 2 $\&$ 3). The results of this contamination–decontamination validation with a controlled contamination model are shown in Appendix \ref{app:cont_decont_validation}. 

The \figreff{fig:contamination_from_mock_map_alpha1_ratio1} in Appendix \ref{app:cont_decont_validation} presents the $\alpha=1$ case, where the mitigation strategy using \namaster successfully recovers the input $C_\ell^{gg}$ despite strong large-scale contamination, since the contamination is linear. We test \regressis methods, giving similar results as \namaster. Similarly, the cross-spectrum $C_\ell^{\kappa g}$ exhibits smaller contamination and is well-behaved, before and after correction. 

 Then, \figreff{fig:contamination_from_mock_map_alpha3_ratio1} in Appendix \ref{app:cont_decont_validation} illustrates the $\alpha=3$ case, introducing non-linear contamination. More interestingly, in this scenario, only the RF \regressis approach efficiently recovers the input spectrum, whereas the linear methods (either \regressis or \namaster) fail. The effect on $C_\ell^{\kappa g}$ remains minimal in both cases (before and after contamination).  From these examples, we see that \namaster reliably recovers the input $C_\ell^{gg}$ when the dependence is linear, which is rather simplistic regarding the trend observed in the UNIONS data (see \figreff{fig:overdensity_VS_template_namaster}). Fortunately, recovering the cross-spectrum $C_\ell^{\kappa g}$ is largely unaffected by these procedures, emphasizing its robustness to tracers' angular systematics.  

Finally, we note that the contamination–decontamination is impacted when the original $C_\ell^{gg}$ used to extract the RF \regressis weight map (in Step 1) differs significantly from that of the contaminated mocks (in Step 2). In the simplest linear contamination case ($\alpha=1$), \figreff{fig:contamination_from_mock_map_alpha1_ratio10} in Appendix \ref{app:cont_decont_validation} shows the corrected spectra when the original $C_\ell^{gg}$ (resp. $C_\ell^{\kappa g}$) is 10 times (resp. 5 times) larger than the mocks\footnote{e.g., corresponding to a galaxy bias to be five times higher for the original mock than the successive other mocks}. Even in this scenario, both linear and non-linear methods recover the true $C_\ell^{gg}$, while $C_\ell^{\kappa g}$ remains largely unaffected. We note that the RF \regressis weight map inherently inherits some imprint of the clustering statistics of the original map, which can affect the decontamination if the underlying $C_\ell^{gg}$ differs from the mocks. This demonstrates that cross-spectra are robust against variations in contamination amplitude, whereas auto-spectra are more sensitive. This section also highlights the challenges in validating contamination–decontamination procedures for auto-spectrum, making cross-spectrum a safer measurement when imaging systematics are present in the data.
\subsubsection{Contaminating mocks with UNIONS LBG \regressis weights}
\begin{figure*}
    \centering
    \includegraphics[width=1\linewidth]{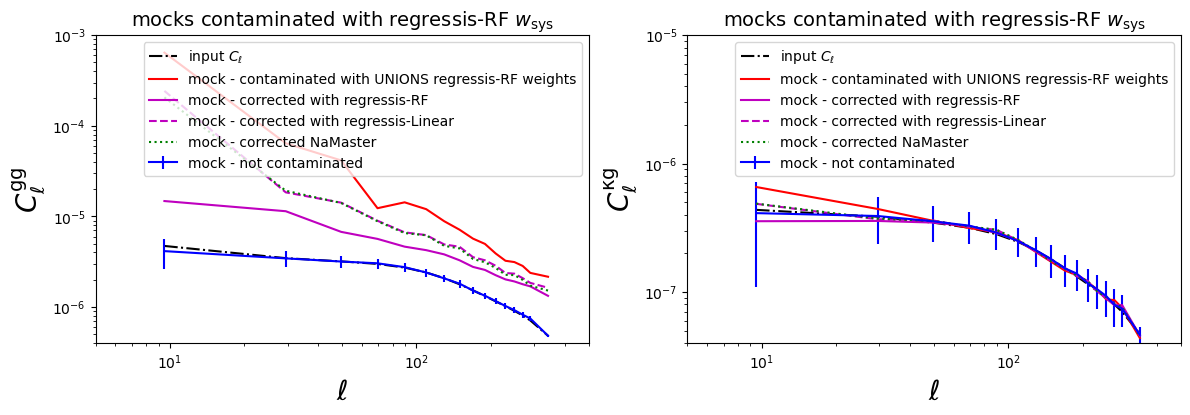}
    \caption{\textbf{Left:} Angular power spectrum of a mock LBG density map, shown before and after contamination using the UNIONS LBG \regressis Random Forest (RF) weight map, and before and after mitigation with \namaster and \regressis.  \textbf{Right:} Same as left, but for the cross-correlation spectrum between the mock LBG density maps and CMB lensing maps.}
    \label{fig:mock_contaminated_with_data_weights}
\end{figure*}

Now that we have validated the contamination–decontamination procedure with a controlled contamination model \citep{Awan2025lsst}, we apply the RF \regressis weights extracted from the actual data to assess the data-based systematic map $w_{\rm sys}$.  The \figreff{fig:wRF_map_diagnostic} and \figreff{fig:wRF_diagnostic} summarize the systematic weights in bins of UNIONS imaging templates. Large corrections, up to $w_{\rm sys} \sim 10$, are required for HEALPix pixels corresponding to very shallow or very deep survey regions in the $u$, $g$, or $r$ bands, as well as for pixels with poor seeing, motivating the removal of “bad” imaging pixels from the UNIONS footprint. Although we do not adopt the RF \regressis weight map as our baseline mitigation method—since it is known to remove significant large-scale power \citep{Krolewski2024fnlcmblensing,Chaussidon2024fnl}—it effectively de-correlates the LBG map, similar to the \namaster outcomes in \figreff{fig:overdensity_VS_template_namaster}. \figreff{fig:wRF_diagnostic_per_template} shows that applying the RF weights removes the positive and negative trends between the observed LBG density and the imaging templates, as expected.
The result of applying these data-driven weights on mock LBG density maps and correcting them with \namaster (and \regressis) is shown in \figreff{fig:mock_contaminated_with_data_weights}. 

When applying the \regressis RF weights derived from UNIONS LBG data to mocks and looking at the mock LBG auto-spectrum, we observe the same behavior as in the mock tests, when the original $C_\ell^{gg}$ and the mock $C_\ell^{gg}$ are different: neither linear nor non-linear methods fully recover the input LBG auto-spectrum, and this is likely because our theoretical model $C_\ell^{gg}$ used for the mocks differs from the underlying UNIONS LBG one. After increasing the amplitude of the mock $C_\ell^{gg}$ (not shown here), we found that the discrepancy between the input and RF-\regressis corrected auto-spectra decreases, but the auto-spectrum remains highly sensitive to these assumptions. In contrast, the cross-spectrum of the mock LBG with external datasets is relatively robust to the mitigation strategy and to uncertainties in the underlying LBG clustering. Consequently, we have conducted these tests so that we focus exclusively on cross-correlations in the next section, as a reliable statistical interpretation of the LBG auto-spectrum is not currently feasible.

We still observe that angular systematics related to inhomogeneities in depth, seeing, dust reddening, and stellar density increase the variance of the measured LBG angular power spectrum, both in auto-correlation and in cross-correlation with CMB lensing (see \figreff{fig:variance_mock}). Even after applying RF-based corrections, which is expected to perform better than linear correction, part of the residual variance — particularly on large scales — remains linked to the intrinsic variance of potentially uncorrected systematics. This point will be important in the next section.

\section{Cross-correlation UNIONS LBGs with external datasets}
\label{sec:cross-correlating_lbg_ext}
In the preceding section, we investigated the feasibility of measuring the cross-correlation signal between contamination-corrected mock UNIONS LBGs and external datasets — illustrated through the case of \textit{Planck} CMB lensing maps, though the framework naturally extends to other projected fields. This mock-based analysis demonstrated that, even in the presence of (apriori known) large-scale angular systematics, the cross-power spectrum between mock UNIONS LBGs and mock \textit{Planck} CMB lensing maps can be robustly recovered, with imaging systematics primarily manifesting as an increase in the variance of the estimator, both before and after correction.

We now turn to the measurement of the cross-correlation signal between the actual UNIONS LBG catalog and external datasets. Specifically, we first cross-correlate UNIONS LBGs with the \textit{Planck} CMB lensing convergence map, and subsequently with quasar datasets from DESI DR1 and \textit{Quaia}. This section should be regarded as a proof of concept rather than a definitive cosmological measurement. Relying on the simplest modeling assumptions and a conservative approach, our primary goal is to establish that the cross-correlation signal of
u-dropout LBGs can be detected over an unprecedentedly large sky footprint — a measurement that, to our knowledge, has not previously been achieved for this galaxy population — thereby paving the way for more complete and precise cosmological analyses in the future.

\subsection{With \textit{Planck} CMB lensing map}

The CMB photons are gravitationally lensed by the intervening large-scale structure along the line of sight between the last-scattering surface and the observer. This affects the observed primary anisotropies and imprints characteristic non-Gaussian signatures in the observed CMB maps~\citep{Lewis2006cmblensing}. Quadratic estimators~\citep{Okamoto2003cmblensing} exploit this effect to reconstruct the associated lensing convergence field, which encodes the weighted projected matter distribution along the line-of-sight from $z=1100$ to $z=0$. In the following, we make use of the \textit{Planck} PR4 CMB lensing map derived from the Planck PR4 temperature and polarization data~\citep{Carron2022PR4Plancklensing}, which represents the state-of-the-art full-sky lensing reconstruction from \textit{Planck} data.

\begin{figure*}
    \centering
    \includegraphics[width=0.47\linewidth]{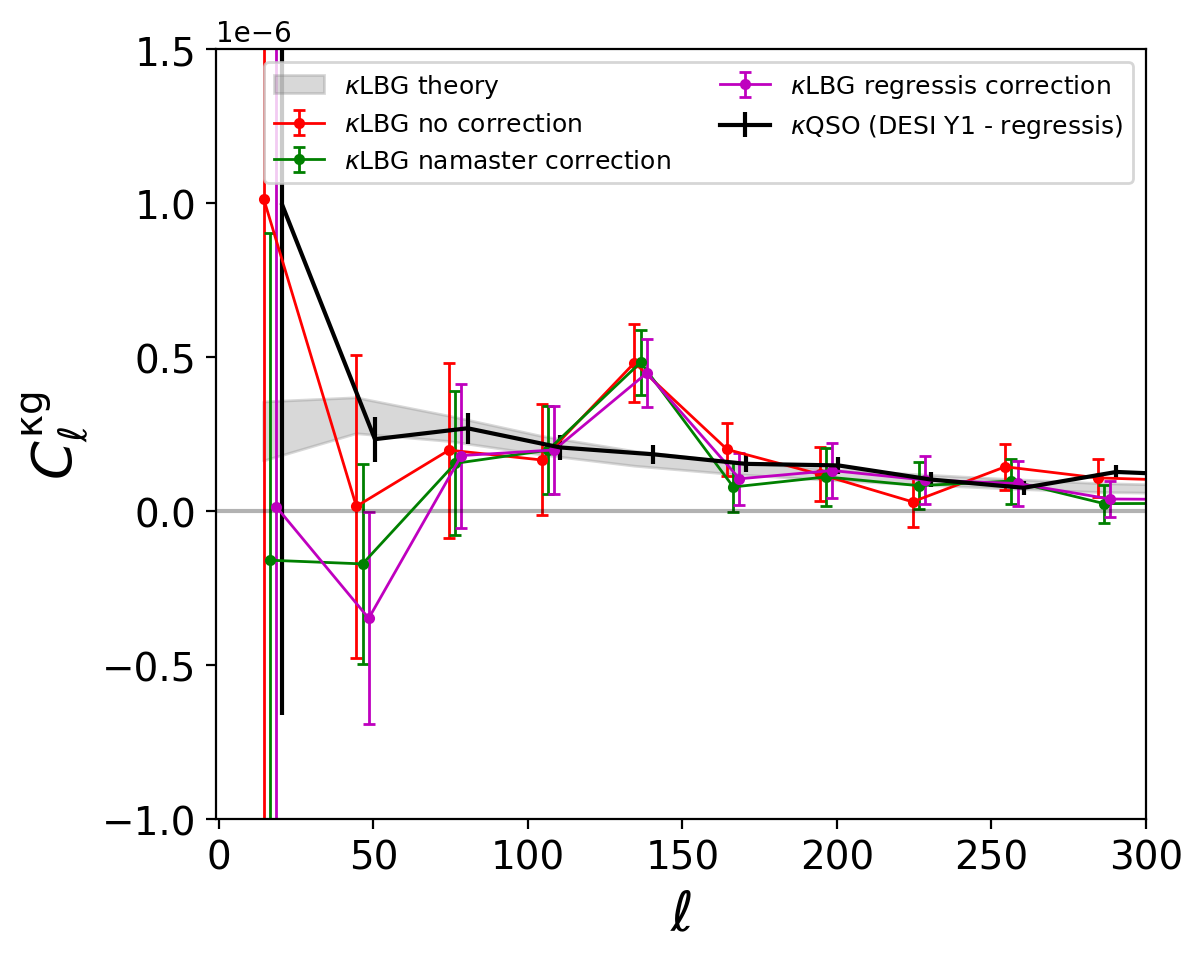}
    \includegraphics[width=0.47\linewidth]{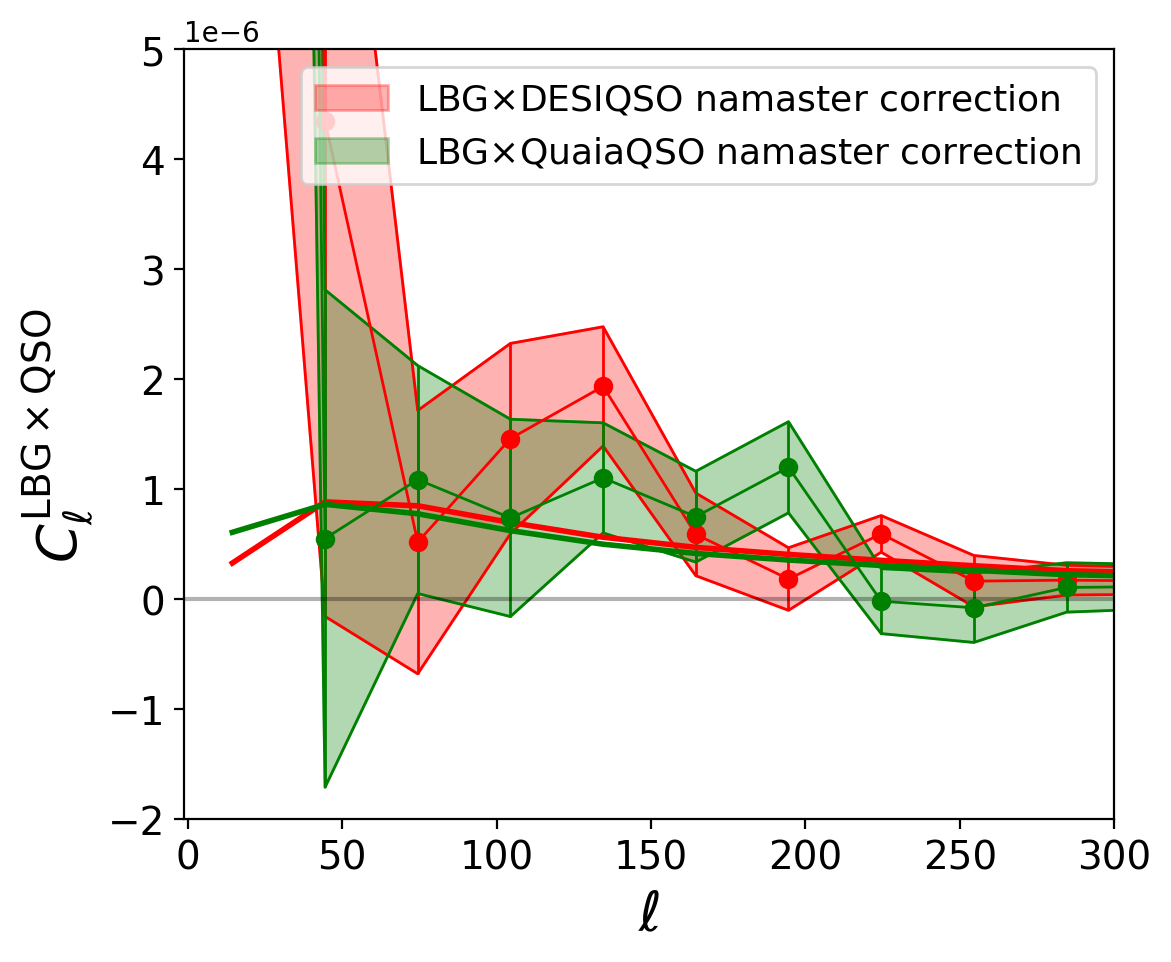}
    \caption{Left: Angular power spectrum between UNIONS LBGs (or DESI DR1 QSOs) and the \textit{Planck} PR4 CMB lensing map. Right: Angular power spectrum between the UNIONS LBGs and the DESI DR1 QSO sample, as well as the \textit{Quaia} sample.}
    \label{fig:LBGkappa_LBGQSO_data}
\end{figure*}

Over the sky region jointly covered by the UNIONS LBG footprint and the \textit{Planck} PR4 lensing mask --- corresponding to approximately $6.6\%$ of the full sky --- we show in the left panel of \figreff{fig:LBGkappa_LBGQSO_data} the binned, decoupled cross-power spectrum between the uncorrected UNIONS LBG angular number density field and the \textit{Planck} PR4 CMB lensing convergence map, measured in 10 $\ell$-bins spanning $\ell = 0$ to $\ell = 300$. The first multipole bin ($0 < \ell < 30$) deserves particular caution: the angular scales probed within this bin are comparable to, or larger than, the effective angular extent of the survey footprint, placing us in a regime where mode-coupling and finite-area effects become significant \citep{Hivon2002master}. In this limit, the mode-coupling matrix $\mathcal{M}_{\ell\ell'}$ is poorly conditioned, the decoupling procedure of Eq.~\eqref{eq:cl_b} becomes unreliable, and the recovered power may be significantly contaminated by mask-induced leakage. We nevertheless retain this bin in the figures for completeness, but caution that it should not be used for cosmological inference (we will see later that the associated variance for that bin is significantly larger, making negligible effects on parameter inference anyway). The cross-power spectrum reveals a positive correlation at nearly all scales, albeit with significant noise despite the broad $\ell$-binning scheme. Upon applying the systematics correction to the UNIONS LBG catalog using \namaster, the signal at small scales ($\ell > 100$) appears largely unaffected, while the clustering amplitude is somewhat suppressed at larger scales ($\ell < 100$). A consistent trend is observed when instead applying the alternative Random Forest-based correction from \regressis, lending further support to the robustness of this behavior.

As a consistency check, we overplot the fiducial theoretical prediction for the cross-spectrum, in which we consider (i) the aforementioned UNIONS LBG photometric redshift distribution (obtained on the XMM-LSS field) considered the truth (ii) the LBG linear bias $b(z)$ modeled following \citet{WilsonWhite2019dropout}, with an overall rescaling $b'(z) = b_0 b(z)$, where $b_0 = 0.65$ was found to provide a good description of the LBG bias inferred from clustering measurements for the same LBG selection (see \citealt{Payerne2025lbgforecast}) (iii) a magnification bias from \citet{Payerne2025lbgforecast} and a RSD component. For further comparison, we also overplot the cross-power spectrum between DESI DR1 QSOs and the same CMB lensing map~\citep{Chaussidon2024fnl,Chiarenza2025cmblensing}, measured over approximately $22\%$ of the sky, which provides a useful benchmark as this sample exhibits a similar redshift coverage and large-scale bias amplitude to that of LBGs. 
 
We nonetheless note that, for the UNIONS LBGs, the fifth $\ell$-bin exhibits a significant deviation from the theoretical prediction, which may be attributed to residual imaging systematics not fully mitigated by the \namaster or \regressis corrections, or to mode-coupling effects arising from the complex geometry of the survey footprint. We emphasize, however, that this first measurement demonstrates that the cross-correlation signal between UNIONS LBGs and the \textit{Planck} PR4 CMB lensing map is broadly consistent with cosmological expectations in terms of amplitude. 

To compare the amplitude of the UNIONS LBG-CMB lensing spectrum with a reference, we choose to use a QSO dataset, covering roughly the same redshift range, with comparable bias. For that, we use in this section the DESI DR1 QSO sample that we cross-correlate with the \textit{Planck} CMB lensing map; the DESI DR1 QSO sample~\citep{Chaussidon2023TSQSO} spans the redshift range $1.6 < z < 3.5$ with an average surface density of $180\,\rm{deg}^{-2}$ over 22 $\%$ of the sky (compared to $6\%$ for the UNIONS LBG sample). Target selection relies on three optical bands ($g, r, z$) from the DESI Legacy Imaging Surveys and two infrared bands ($W_1, W_2$) from the Wide-field Infrared Survey Explorer (WISE) and was extensively validated during the DESI Survey Validation campaign. The final quasar target selection is based on a Random Forest algorithm, selecting objects in the magnitude range $16.5 < r < 23$. Spectroscopic redshifts are assigned to each confirmed quasar using three complementary algorithms: the DESI pipeline classifier \texttt{Redrock}, a broad Mg\,\textsc{ii} line finder, and the machine learning-based classifier \texttt{QuasarNET}. Each object is further assigned a weight to account for spectroscopic redshift failures, fiber assignment incompleteness, and imaging systematics, the latter being corrected via a linear regression approach implemented in \regressis (see~\citealt{Chaussidon2024fnl} for details). The DESI QSO density field is therefore expected to be free of large-scale imaging systematics at the scales of interest~\citep{Chaussidon2024fnl}.
 and compares favorably with the analogous signal measured using DESI DR1 QSOs.

For any cosmological inference from this signal, an accurate characterization of the associated covariance is essential. The Gaussian-only covariance of the binned decoupled angular power spectrum at a given multipole $\ell$ can be expressed as a function of the respective auto-, cross-, and noise power spectra (\citealt{Brown2005cellcov}, see Appendix~\ref{sec:ang_pow_cov}), and can therefore be computed analytically from the fiducial power spectra and input noise levels. The corresponding Gaussian-only error bars are shown as the shaded region in the left panel of \figreff{fig:LBGkappa_LBGQSO_data}. We see by eye that the error bars seem to be small compared to the observed scatter of the data points. To check whether the Gaussian-only approximation alone is sufficient to capture the full variance of the estimator, we compute the delete-one Jackknife covariance matrix \citep{tukey1958} for the same set of maps. This method estimates the variance of the estimator by successively removing one patch at a time from a partition of $N_{\rm JK}$ regions covering the survey footprint, remeasuring the cross-power spectrum for each realization, and assembling the covariance from the $N_{\rm JK}$ resulting spectra with the appropriate $(N_{\rm JK}-1)/N_{\rm JK}$ prefactor. The UNIONS LBG footprint is divided into $N_{\rm JK}=110$ Jackknife regions (see \figreff{fig:variance_mock}). In the absence of imaging systematics, the Jackknife covariance is expected to recover the Gaussian-only variance for both auto- and cross-spectra, as validated on mock maps in Appendix~\ref{app:jackknife_validation_with_mock}. The resulting Jackknife error bars estimated on the real UNIONS LBG data are shown on each data point in the left panel of \figreff{fig:LBGkappa_LBGQSO_data}, and are clearly in exceeding of the Gaussian-only variance (the ratio between the Jackknife and Gaussian-only variances reaches a factor of $\sim 3$ at small scales, and between $10$ and $15$ at large scales, see the left panel of \figreff{fig:LBGkappa_mcmc}). This excess variance is likely attributable to residual angular systematics that are not fully mitigated by our regression-based correction procedures. 

\subsection{Cross-correlation with \textit{Planck} CMB lensing: impact on $\fnl$}
 
This excess variance in the cross-power spectrum between UNIONS LBGs and the \textit{Planck} PR4 CMB lensing map, driven by residual angular systematics, directly impacts cosmological inference by significantly inflating the uncertainties on the recovered parameters. This is particularly consequential for the measurement of local primordial non-Gaussianity (PNG), one of the primary science objectives of LBG samples from Stage-4 photometric surveys such as the Rubin LSST \citep{LSST2009whitepaper}, as well as from spectroscopic surveys including DESI-II \citep{Schlegel2022DESI2} and Stage-5 facilities such as WST \citep{Mainieri2024wst} or Spec-S5 \citep{Besuner2025specs5}. Local PNGs, parameterized by $\fnl$, induces a scale-dependent bias correction \citep{Dalal2008,Slosar2008,Desjacques2010fNL} (see Appendix \ref{sec:modeling_ang_pow_spec})
\begin{equation}
    \Delta b(z,k) = b_\Phi \fnl \frac{3 \Omega_{\rm m} H_0^2}{2 k^2 T(k) D(z)},
\end{equation}
with $b_\Phi = 2 (b(z) - p_\Phi)\delta_c$, where $b(z)$ is the linear galaxy bias (in this work, we use the bias prescription proposed by \citet{WilsonWhite2019dropout}, evaluated at the detection limit $m_r=24.2$ of the LBG sample), $T(k)$ is the transfer function. The parameter $p_\Phi$ depends on the tracer population \citep{Slosar2008,Barreira2020a} (in this work, we use the universality relation, giving $p_\Phi=1$). It is a large-scale effect $\sim 1/k^2$, thus requiring precise measurement of the large-scale clustering power, enabled by high-redshift tracers such as LBGs, probing large volumes. 
 
\begin{figure*}
    \centering
    \includegraphics[width=0.45\linewidth]{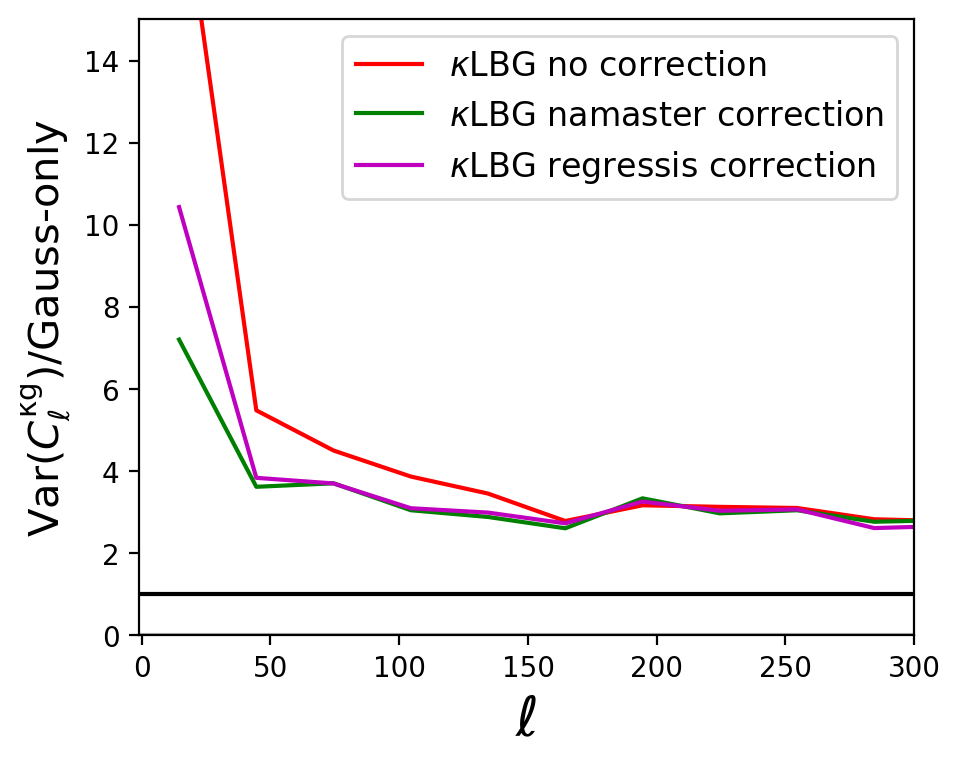}
    \includegraphics[width=0.42\linewidth]{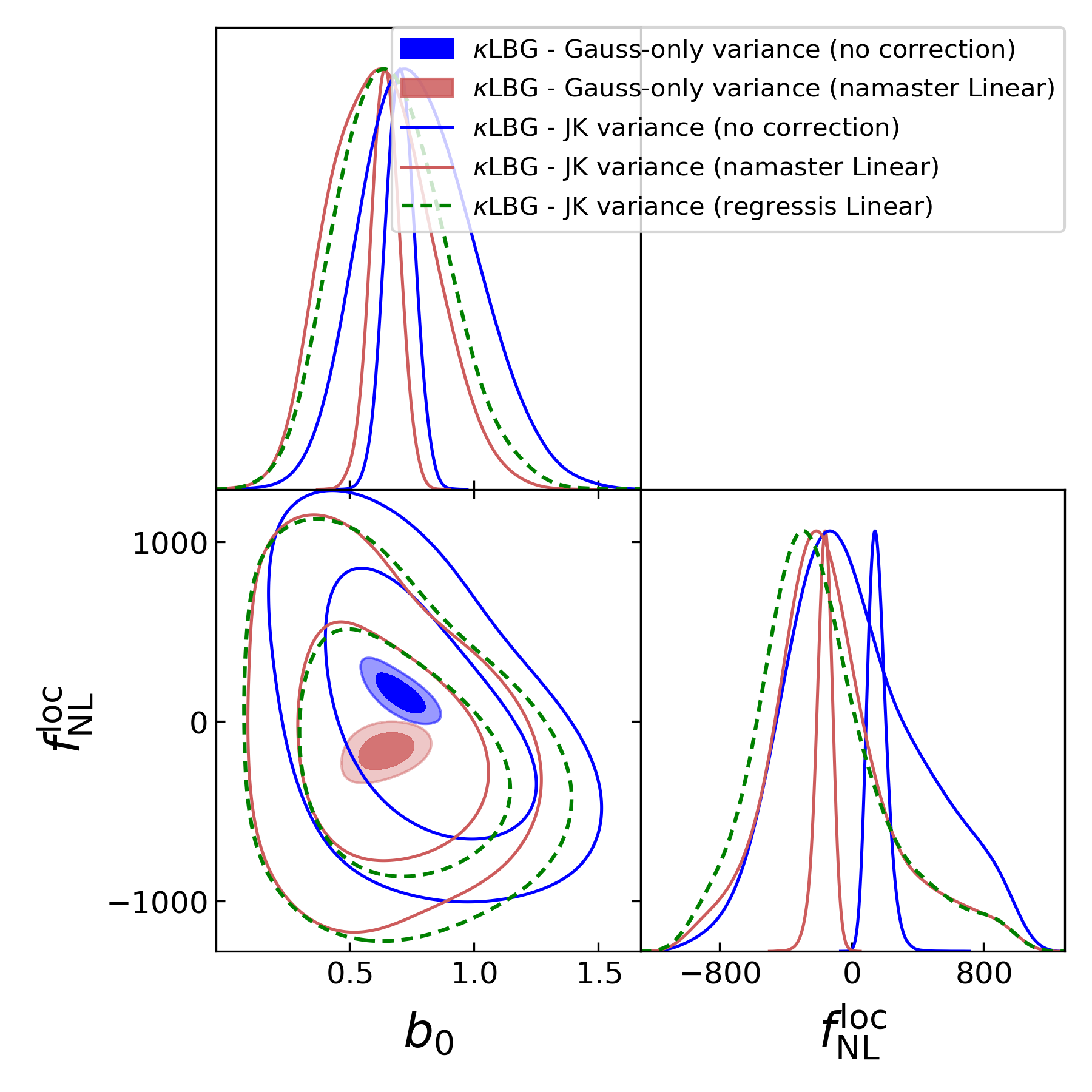}
    \caption{Left: Ratio of the diagonal elements of the Jackknife covariance matrix to those of the Gaussian-only covariance matrix. Right: Two-dimensional posterior distribution of $\fnl$ and $b_0$ inferred from the $C_\ell^{\kappa g}$ measurements, using either the theoretical or Jackknife covariance.}
    \label{fig:LBGkappa_mcmc}
\end{figure*}

This scale-dependent bias effect is detectable both on the 3D and \textit{projected} tracer power spectrum (either auto-spectrum or cross-spectrum with external tracer). Thus, we can evaluate the constraining power of the UNIONS LBG sample on local PNG by fitting the cross-power spectrum with the \textit{Planck} PR4 CMB lensing map.\footnote{The $\kappa$LBG transfer function is estimated from uncontaminated mocks, by taking the ratio of the cross-spectrum of the corrected mocks to that of the uncorrected ones. The transfer function is about $80\%$ for the first $\ell$-bin, and 1 elsewhere. We do not account for de-projection bias, as it is several orders of magnitude smaller than the signal.} Adopting a Gaussian likelihood, we report in \tabreff{tab:fnl_fit_LBG_kappa} the posterior means and standard deviations of $f_{\rm NL}^{\rm loc}$ and the bias amplitude $b_0$ --- which rescales the bias prescription of \citet{WilsonWhite2019dropout} --- together with the two-dimensional posterior distributions shown in the right panel of \figreff{fig:LBGkappa_mcmc}.
 
When adopting the Gaussian-only covariance in the Bayesian inference, the recovered uncertainty on $f_{\rm NL}^{\rm loc}$ is of order $\sigma(f_{\rm NL}^{\rm loc}) \sim 50$. Upon replacing the Gaussian-only covariance with the Jackknife estimate, this uncertainty increases dramatically to $\sigma(f_{\rm NL}^{\rm loc}) \sim 350$, a direct consequence of the excess variance at large angular scales where the sensitivity to $f_{\rm NL}^{\rm loc}$ is greatest. The constraints on the bias amplitude $b_0$ are consistent with the value inferred from clustering redshifts in \citep{Payerne2025lbgforecast}, with a factor of $\sim 2$ degradation in precision between the Gaussian-only and Jackknife covariances --- as expected, since the bias is primarily constrained by the small-scale signal where the influence of $f_{\rm NL}^{\rm loc}$ is subdominant.
 
This exercise illustrates the critical impact of imaging systematics on the variance of the $\kappa$LBG cross-power spectrum and, consequently, on the constraining power on $f_{\rm NL}^{\rm loc}$: residual systematics inflate the scatter of the estimator preferentially at large scales, precisely where the $f_{\rm NL}^{\rm loc}$ signal-to-noise is highest, making their mitigation a key requirement for future cosmological analyses.
\begin{table}[]
    \centering
    \resizebox{0.48\textwidth}{!}{\begin{tabular}{c|c|c}
    Study cases&$\fnl\pm\sigma(\fnl)$&$b_0\pm\sigma(b_0)$\\
    \hline
Gauss-only variance (no correction)& $158  \pm  56$  &  $0.7  \pm  0.1$ \\ 
Gauss-only variance (\namaster Linear)& $-166  \pm  49$  &  $0.6  \pm  0.1$ \\ 
JK variance (no correction)& $27  \pm  401$  &  $0.8  \pm  0.2$ \\ 
JK variance (\namaster Linear) &$-130  \pm  361$  &  $0.6  \pm  0.2$ \\ 
JK variance (\regressis Linear)& $-189  \pm  377$  &  $0.7  \pm  0.2$ \\ 
    \end{tabular}}
    \caption{Best-fit values of $\fnl$ with associated uncertainties, together with the bias rescaling factor $b_0$ used in the bias prescription of \citet{WilsonWhite2019dropout}.}
    \label{tab:fnl_fit_LBG_kappa}
\end{table}

\subsection{With QSO datasets: DESI DR1 and \textit{Quaia}}
Complementary to the \textit{Planck} CMB lensing map, the redshift coverage of the UNIONS LBG sample overlaps significantly with external QSO datasets, namely the DESI DR1 QSO catalog~\citep{Chaussidon2024fnl} ($3\%$ of the sky in common with the UNIONS LBG sample) --- which we also employ as an independent validation of the LBG--CMB lensing cross-correlation measurement --- and the \textit{Quaia} catalog~\citep{Fisher2024quaia} ($6\%$ of the sky in common between \textit{Quaia} and the UNIONS LBG sample).
 
The \textit{Quaia} catalog~\citep{Fisher2024quaia} was constructed by cross-matching quasar candidates from the third \textit{Gaia} data release with photometry from the unWISE reprocessing of WISE, improving the rejection of stellar contaminants. We make use of the bright sample ($G < 20.0$), which is less sensitive to sky background contamination and photometric redshift misestimation than the full \textit{Quaia} sample ($G < 20.5$). The sample spans the redshift range $0 < z < 3.5$ with an average surface density of $25\,\rm{deg}^{-2}$. Redshifts in the \textit{Quaia} catalog are estimated from \textit{Gaia} and unWISE photometry, supplemented by \textit{Gaia}-based spectral photometry.
 
Accounting for the selection function of each QSO dataset, we show in the right panel of \figreff{fig:LBGkappa_LBGQSO_data} the cross-power spectra between the UNIONS LBG sample and each of the two QSO datasets, corresponding to approximately $5-6\%$ of the full sky (when combined). The theoretical predictions\footnote{Computed accounting for each tracer's redshift distribution (photometric for UNIONS LBGs, spectroscopic for DESI and \textit{Quaia}), magnification bias, and bias prescriptions from \citep{Alonso2023quaia} for \textit{Quaia} and \citep{Chaussidon2024fnl} for DESI DR1 QSOs.} are in overall good agreement with the measurements, despite the large scatter, as well as the larger clustering amplitude measured in the first bin, for which it is necessary to be cautious due to the scales probed relative to the size of the common footprint between UNIONS LBG and QSO datasets. This demonstrates that the clustering amplitude is recovered within a reasonable range and confirms that the cross-correlation signal between UNIONS LBGs and QSO tracers is detected at the expected level, while further work will be needed to fully exploit all angular scales in a consistent cosmological framework.
\section{Conclusions}

In this work, we have investigated the feasibility of exploiting UNIONS-selected LBGs as cosmological tracers of large-scale structure at redshift $z \gtrsim 2$. Based on a $u$-dropout selection applied to the UNIONS GAaP $ugri$ catalog~\citep{Gwyn2025unions}, we have characterised the angular clustering properties of this sample and assessed the potential of cross-correlation techniques, with particular emphasis on CMB lensing and quasar datasets.
 
A central finding of this study is that, while UNIONS represents a promising dataset for high-redshift galaxy clustering, the LBG auto-angular power spectrum is severely affected by spatially varying imaging systematics. Variations in PSF depth, seeing~\citep{Gwyn2025unions}, Galactic extinction $E(B-V)$~\citep{Schlegel1998ebv}, and stellar density~\citep{Gaia2023stardens} introduce large-scale fluctuations in the observed number density that cannot be fully captured by standard mitigation schemes alone. Even after applying state-of-the-art correction strategies --- including template regression and de-projection techniques implemented in \namaster --- residual contamination persists on the largest angular scales, suggesting either that the adopted methodology does not fully capture the spatial structure of the systematics, or that additional sources of angular contamination beyond those considered here are affecting the data. Consequently, the corrected auto-spectrum exhibits significant deviations from theoretical expectations, preventing a reliable cosmological interpretation without further methodological developments or more stringent control of large-scale survey systematics. It is worth noting that the challenges encountered here are not specific to UNIONS, but are generic to any wide-field photometric survey operating near its detection limit with a faint tracer population.
 
In contrast, cross-correlation measurements provide a substantially more robust avenue for cosmological analysis. By cross-correlating the UNIONS LBG sample with external datasets --- including the \textit{Planck} PR4 CMB lensing convergence map~\citep{Carron2022PR4Plancklensing}, DESI DR1 quasars~\citep{Chaussidon2023TSQSO}, and the \textit{Quaia} catalog~\citep{Fisher2024quaia} --- we consistently recover a positive clustering signal across a broad range of angular scales. This robustness stems from the fact that the dominant observational contaminants are uncorrelated between the independent datasets, rendering the cross-correlation largely insensitive to imaging systematics. Mock-based validation further supports this conclusion, demonstrating that even in the presence of strong, non-linear, and spatially varying systematics, the cross-power spectrum can be reliably recovered.
 
We note, however, that residual systematics continue to play a non-negligible role in shaping the statistical uncertainties of the cross-correlation measurements. In particular, Jackknife covariance estimates reveal significant excess variance at large angular scales, precisely where the cosmological signal is most sensitive to local primordial non-Gaussianity~\citep{Slosar2008}. As illustrated through a simplified $f_{\rm NL}^{\rm loc}$ analysis, this excess variance directly degrades the constraining power on local PNG, with constraints weakening considerably when the full empirical covariance is employed. This underscores the importance of accurately characterizing survey systematics not only at the level of the mean signal but also in the covariance of clustering estimators.
 
Despite these limitations, the cross-correlation signal between UNIONS LBGs and external tracers is detected at an amplitude consistent with fiducial bias and redshift distribution models. The agreement with both the \textit{Planck} CMB lensing and DESI QSO cross-correlations supports the interpretation that the selected LBG population traces the underlying matter distribution in a physically consistent manner, probing similar redshift and bias regimes. This constitutes, to our knowledge, one of the first statistically significant detections of LBG cross-correlations over a large sky area at $z \sim 2.5$.
 
Taken together, our results establish that UNIONS-selected LBGs can be used to trace high-redshift large-scale structure, while highlighting the challenges inherent to photometric LBG clustering analyses on large angular scales. Addressing these challenges will require a combination of deeper control of imaging systematics, more sophisticated non-linear mitigation techniques, and refined modeling of selection effects.
 
Looking ahead, upcoming wide-field photometric surveys such as the Rubin LSST~\citep{LSST2009whitepaper} and next-generation spectroscopic facilities including DESI-II~\citep{Schlegel2022DESI2} and the WST~\citep{Mainieri2024wst} will dramatically expand both the statistical power and sky coverage available for LBG studies. In this context, cross-correlations with CMB lensing and spectroscopic tracers will remain an indispensable tool for robust cosmological inference, offering a clear pathway toward competitive constraints on local PNG and the growth of structure across the high-redshift Universe.

\section*{Acknowledgments}

We thank the developers and maintainers of the following software tools used in this work:
\texttt{NumPy}\footnote{\url{https://numpy.org}} \citep{vanderWaltnumpy}, \texttt{HEALPix}\footnote{\url{https://healpix.sourceforge.io}} \citep{healpix2004}, \texttt{SciPy}\footnote{\url{https://scipy.org}} \citep{jonesscipy}, \texttt{Matplotlib}\footnote{\url{https://matplotlib.org}} \citep{Hunter2007matplotlib}, \texttt{GetDist}\footnote{\url{https://getdist.readthedocs.io}} \citep{Lewis2025getdist}, \texttt{emcee}\footnote{\url{https://emcee.readthedocs.io}} \citep{Foreman_Mackey_2013emcee}, \texttt{Jupyter}\footnote{\url{https://docs.jupyter.org}} \citep{jupyter}. Language editing assistance for parts of the manuscript text was provided by Claude (Anthropic). All scientific content, analysis, and conclusions are the sole responsibility of the authors.

We thank David Alonso and Thomas Cornish for their help in using \texttt{NaMaster}. We thank Edmond Chaussidon, Alex Krolewski, and Sofia Chiarenza for their help in using \texttt{regressis} and for useful discussions around the DESI DR1 QSO dataset. Constantin Payerne and Christophe Yèche acknowledge support from grant ANR-22-CE31- 0009 for the HZ-3D-MAP project and from grant ANR-22-CE92-0037 for the DESI-Lya project. William d’Assignies Doumerg acknowledges support from the MICINN projects PID2019-111317GB-C32, PID2022-141079NB-C32, as well as the predoctoral program AGAUR-FI ajuts (2024 FI-1 00692) Joan Oró. IFAE is partially funded by the CERCA program of the Generalitat de Catalunya. Michael J. Hudson acknowledges support from NSERC through a Discovery Grant. 

\bibliographystyle{yahapj}
\bibliography{main}
\begin{appendix}
\section{Formalism: Angular Power Spectrum}
\label{app:formalism}
\subsection{Modeling the agular power spectrum}
\label{sec:modeling_ang_pow_spec}
For a point-like tracer,
\begin{equation}
    q_{\rm g}^{\rm int}(\chi) = b(z)\, n(z)\, \frac{dz}{d\chi},
\end{equation}
where $b(z)$ is the linear galaxy bias and $n(z)$ the normalized redshift distribution. We restrict the analysis to $\ell<300$, corresponding to $k \lesssim 0.05$--$0.075\,h\,\mathrm{Mpc}^{-1}$ at $z \sim 1.5$--3, i.e. well within the linear regime \citep{Takahashi2012kNL}, where the bias is approximately scale-independent \citep{Desjacques2018} (in the absence of local primordial non-Gaussianities). However, local Primordial non-Gaussianity (PNG), parameterized by $\fnl$, induces a scale-dependent bias correction \citep{Dalal2008,Slosar2008,Desjacques2010fNL}
\begin{equation}
    \Delta b(z,k) = b_\Phi \fnl \frac{3 \Omega_{\rm m} H_0^2}{2 k^2 T(k) D(z)},
\end{equation}
with $b_\Phi = 2 (b - p_\Phi)\delta_c$. The parameter $p_\Phi$ depends on the tracer population \citep{Slosar2008,Barreira2020a}. 

Moreover, lensing magnification and redshift-space distortions (RSD) contribute to the total kernel of a given tracer field, such as
\begin{equation}
    q_{\rm g}^{\rm mag}(\chi) = (5s - 2)\, \frac{3 \Omega_{\rm m} H_0^2}{2} \frac{\chi}{a(\chi)} 
    \int_{\chi}^{\chi_{\rm H}} d\chi' \, n(z(\chi')) \frac{\chi' - \chi}{\chi'},
\end{equation}
\begin{equation}
    q_{\rm g}^{\rm RSD}(\chi) = - H(z)\, n(z)\, \frac{d \ln D}{d \ln a}\, j_\ell''(k\chi),
\end{equation}
where $s$ is the magnification bias \citep{Challinor2011magnification}. The observed galaxy power spectrum is
\begin{equation}
    C_\ell^{gg,\mathrm{obs}} = \sum_{ij} C_\ell^{q_i q_j} + \frac{1}{\bar{n}_{\rm gal}},
\end{equation}
with $q_i \in \{q_{\rm g}^{\rm int}, q_{\rm g}^{\rm mag}, q_{\rm g}^{\rm RSD}\}$. CMB photons are gravitationally lensed by large-scale structure, generating a convergence field $\kappa$ reconstructed from temperature and polarization maps \citep{Okamoto2003cmblensing}. Its kernel is \citep{Lewis2006cmblensing}
\begin{equation}
    q_\kappa(\chi) = \frac{3 H_0^2 \Omega_{\rm m}}{2 c^2} \frac{\chi}{a(\chi)} \left( \frac{\chi_* - \chi}{\chi_*} \right),
\end{equation}
which peaks at $z \sim 2$. The observed spectrum is $C_\ell^{\kappa\kappa,\mathrm{obs}} = C_\ell^{\kappa\kappa} + N_\ell^\kappa$. Cross-correlations with galaxies, $\sum_i C_\ell^{\kappa q_i}$, probe the matter distribution and constrain $\fnl$ via scale-dependent bias.

Finally, under the Limber approximation \citep{Limber1953approx}, valid at high multipoles, the angular power spectrum simplifies to
\begin{equation}
    C_\ell^{XY} \approx \int_0^{\chi_{\rm H}} \frac{d\chi}{\chi^2} \, q_X(\chi)\, q_Y(\chi)\, P\!\left( \frac{\ell + 1/2}{\chi}, z(\chi) \right),
\end{equation}

\subsection{Angular Power Spectrum Covariance}
\label{sec:ang_pow_cov}
The covariance of the binned de-coupled angular power spectrum is given by
\begin{equation}
    \mathrm{Cov}(\widehat{C}_b^{XY}, \widehat{C}_{b'}^{ZW}) = B_{b\ell} \, \mathrm{Cov}(\widehat{C}^{XY}_\ell, \widehat{C}^{ZW}_{\ell'}) \, B_{b'\ell'},
\end{equation}
where $B_{b\ell}$ is the binning matrix, and the covariance of the unbinned spectra is \citep{Brown2005cellcov}
\begin{equation}
    \mathrm{Cov}(\widehat{C}^{XY}_\ell, \widehat{C}^{ZW}_{\ell'}) = \frac{C^{XZ}_\ell \, C^{YW}_\ell + C^{XW}_\ell \, C^{YZ}_\ell}{2\ell + 1}\, \mathcal{M}^{-1}_{\ell\ell'},
    \label{eq:covariance_Gauss_pred}
\end{equation}
where $C^{XZ}_\ell$ are expected values for the angular power spectrum amplitude between field $X$ and  $Z$ (that is generally computed using a cosmological model, evaluated at a fiducial cosmology) and relates to the two-point full-sky angular correlation function between field $X$ and $Z$
\begin{equation}
    w(\theta) = \sum_{\ell} \frac{2\ell + 1}{4\pi} \, C_\ell \, P_\ell(\cos\theta)
\end{equation}
and relates to the full-sky scatter of the overdensity field
\begin{equation}
    \langle \delta^2 \rangle = \sum_{\ell} \frac{2\ell + 1}{4\pi} \, C_\ell
\end{equation}
Not included in the above expression are non-Gaussian contributions to the covariance, such as the super-sample covariance (SSC), which is currently considered to be the dominant non-Gaussian contribution. SSC arises from the non-linear modulation of local observables by long-wavelength density fluctuations.
\section{UNIONS templates}
\label{app:unons_templates}
\begin{figure*}
    \centering
    \includegraphics[width=0.9\linewidth]{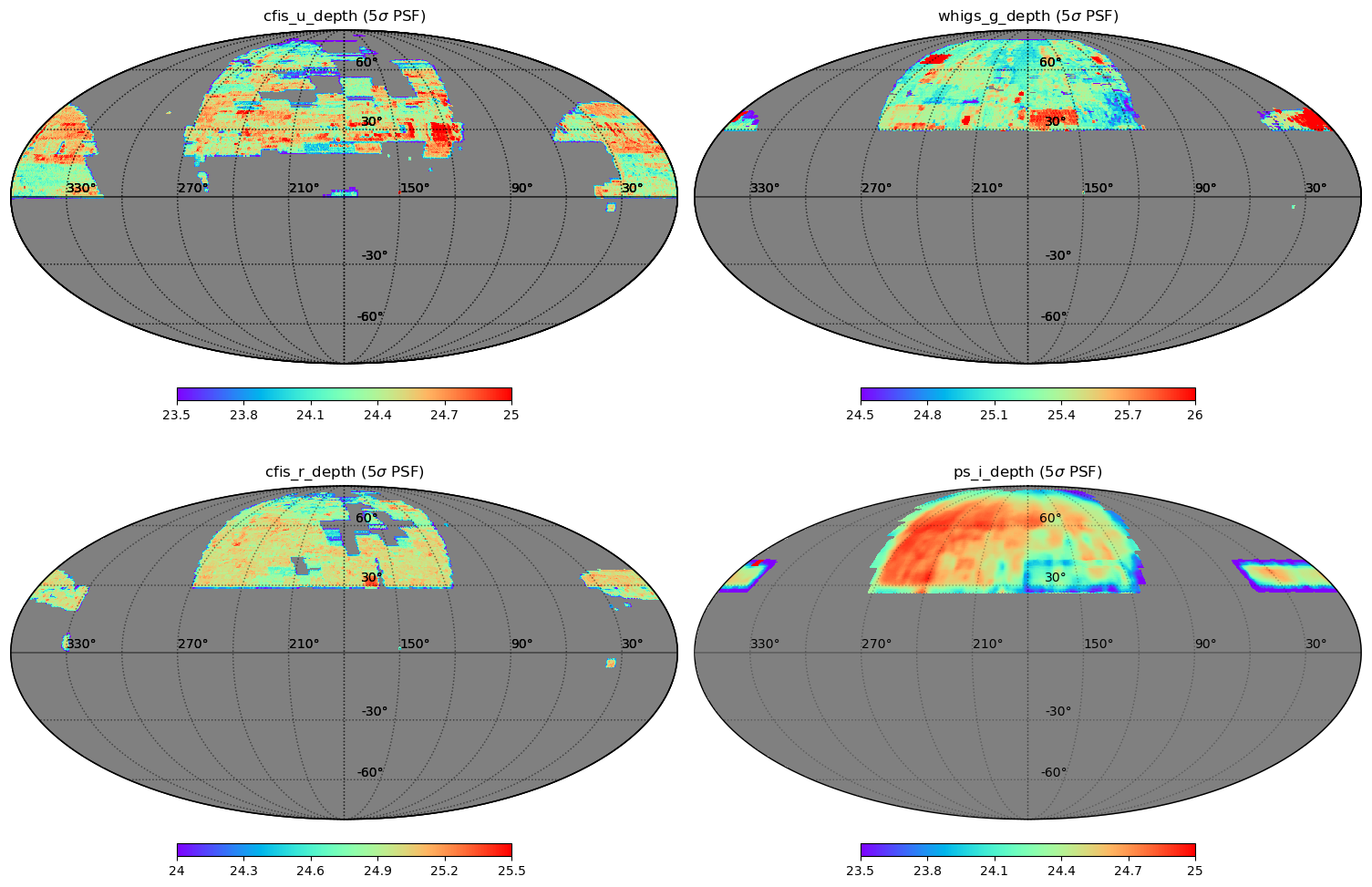}
    \caption{UNIONS PSF depth.}
    \label{fig:unions_depth}

    \centering
    \includegraphics[width=0.9\linewidth]{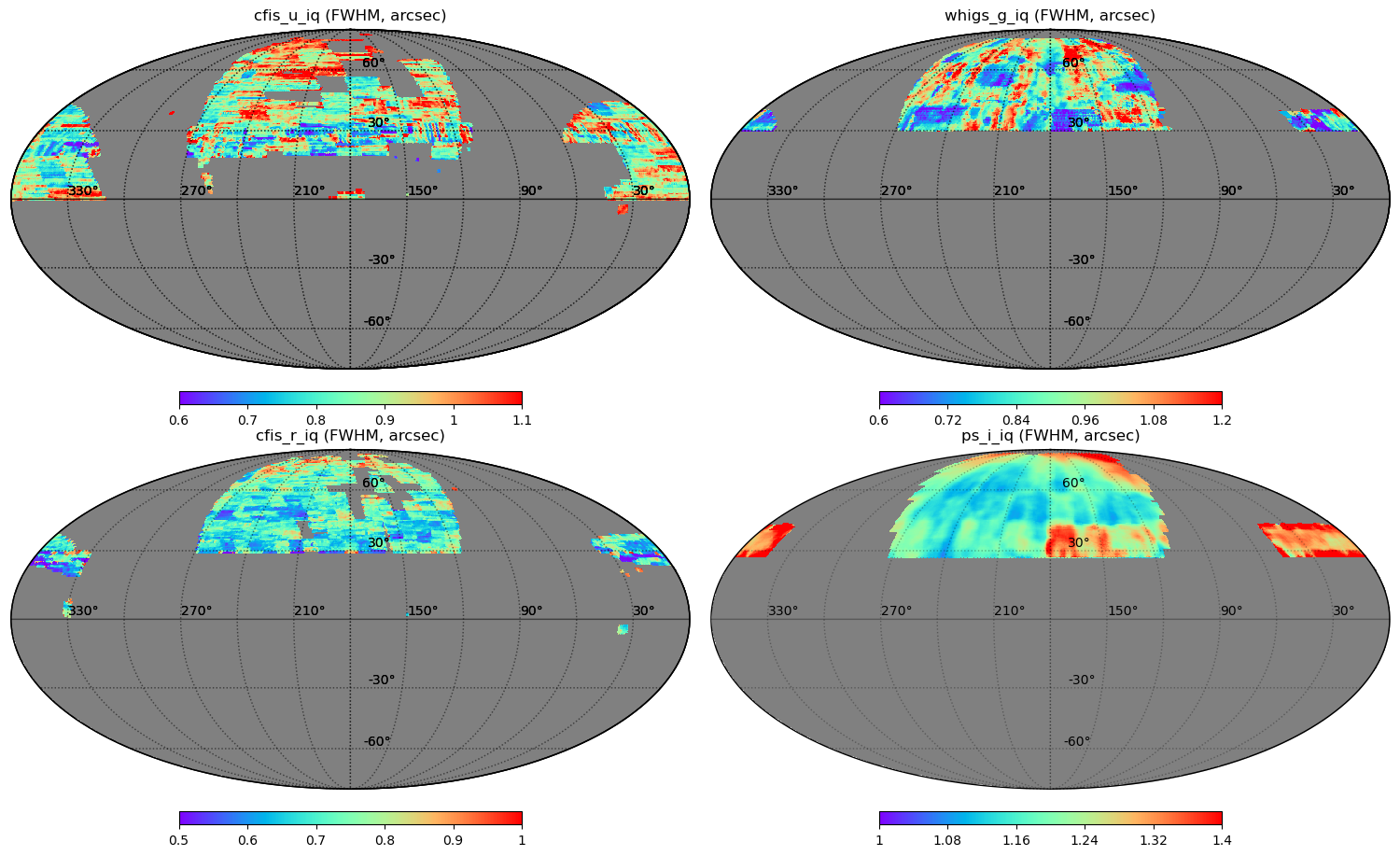}
    \caption{UNIONS image quality.}
    \label{fig:unions_seeing}
\end{figure*}

\begin{figure*}
\centering
        \includegraphics[width=0.9\linewidth]{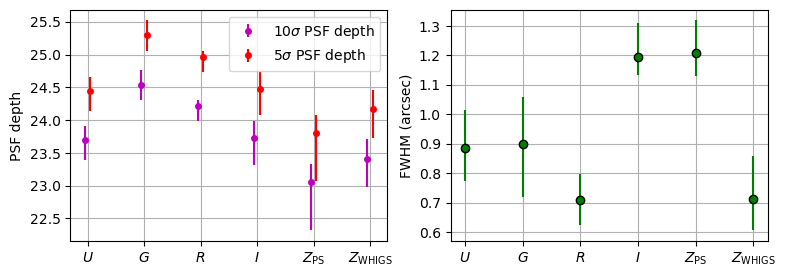}
    \caption{Summary of UNIONS PSF depth and UNIONS image quality.}
    \label{fig:summary_unions}
\end{figure*}

\figreff{fig:unions_depth} shows the PSF depth of different UNIONS sub-surveys/filters. \figreff{fig:unions_seeing} shows the associated image quality. The HEALPix maps are generated based on the map files in \citet{Gwyn2025unions} (see their Figure 3 and 5). \figreff{fig:summary_unions} shows the summary values. 
\section{Contamination-decontamination validation with controlled contamination}
\label{app:cont_decont_validation}
\begin{figure*}
    \centering
    \includegraphics[width=1\linewidth]{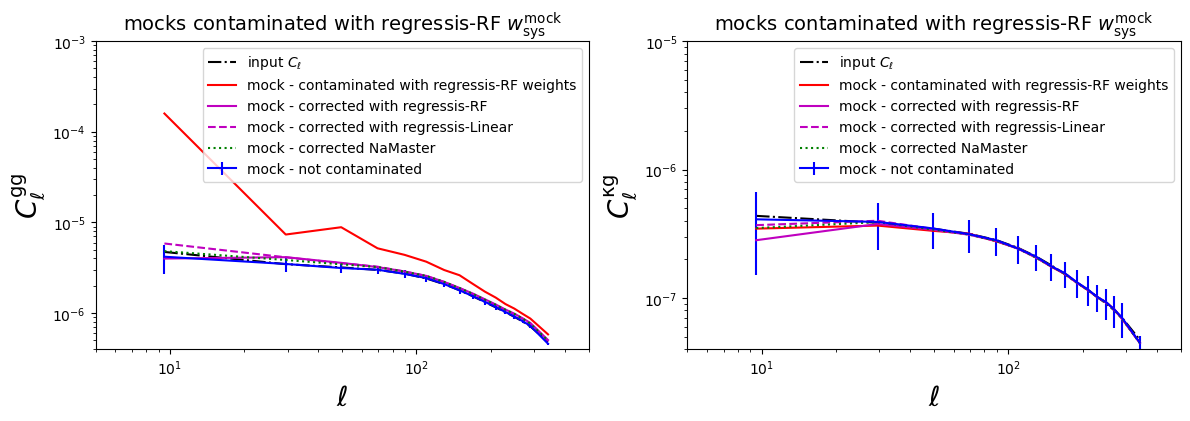}
    \caption{Left: Angular power spectrum of mock LBG density maps, contaminated and decontaminated using a linear contamination model \citep{Awan2025lsst} (i.e., $\alpha=1$ in Eq.~\eqref{eq:linear_contamination_model}), with the same $C_\ell^{\rm gg}$ used for both the original contaminated map and the mocks.  
Right: Same as left, but for the cross-correlation between the mock LBG density maps and the CMB lensing maps.}
    \label{fig:contamination_from_mock_map_alpha1_ratio1}
    \includegraphics[width=1\linewidth]{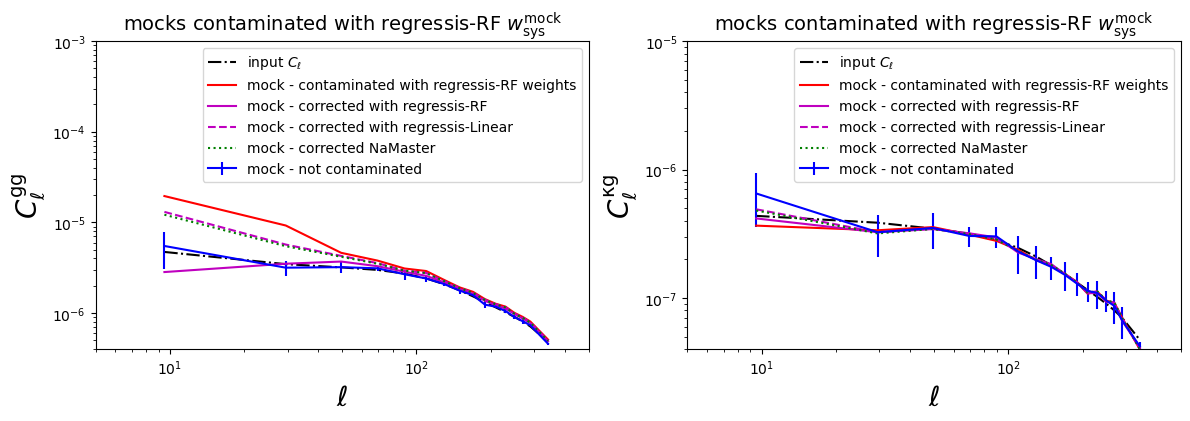}
    \caption{Same as Figure \ref{fig:contamination_from_mock_map_alpha1_ratio1}, but with $\alpha=3$ in Eq.~\eqref{eq:linear_contamination_model}.}
    \label{fig:contamination_from_mock_map_alpha3_ratio1}
    \includegraphics[width=1\linewidth]{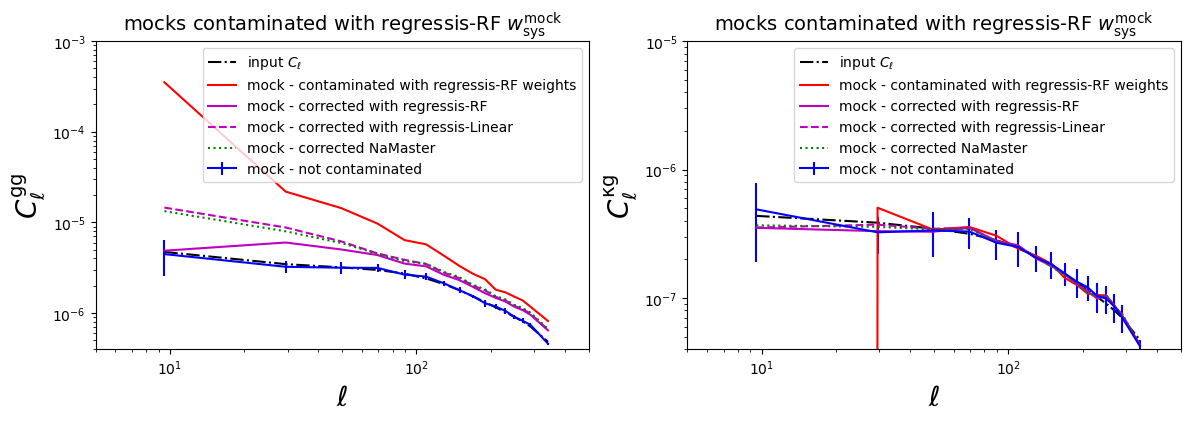}
    \caption{Same as Figure \ref{fig:contamination_from_mock_map_alpha1_ratio1}, but the original $C_\ell^{\rm gg}$ is 10 times larger than the mock $C_\ell^{\rm gg}$.}
    \label{fig:contamination_from_mock_map_alpha1_ratio10}
\end{figure*}

In this appendix, we explore the contamination-decontamination procedure presented in \ref{sec:imag_syst}. Figure \ref{fig:contamination_from_mock_map_alpha1_ratio1} presents the case with mock contamination, with a linear contamination model, Figure \ref{fig:contamination_from_mock_map_alpha3_ratio1} shows the case of a non-linear mock-contamination, and Figure \ref{fig:contamination_from_mock_map_alpha1_ratio10} shows a linear mock contamination, but with a ratio of 10 between the original contaminated mock and the subsequent mocks. 
\begin{figure*}
    \centering
\includegraphics[width=0.45\linewidth]{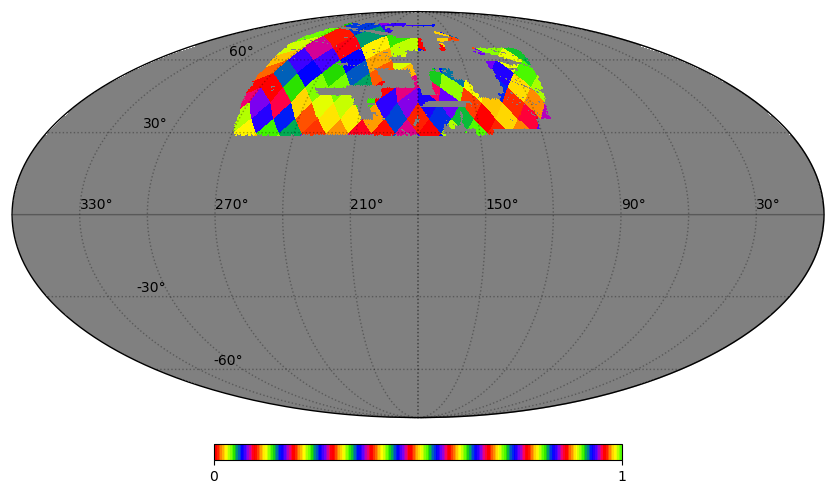}
\includegraphics[width=0.45\linewidth]{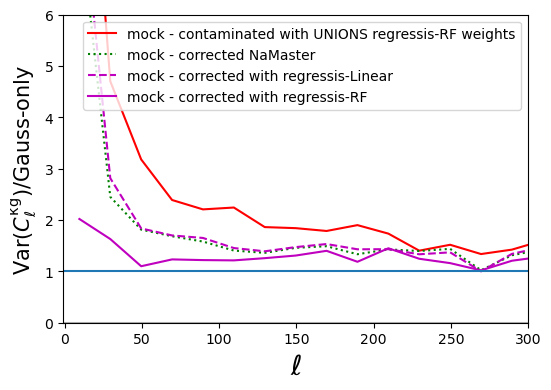}
    \caption{Left: Jackknife binning scheme of the default UNIONS LBG footprint. Right: Ratio between (i) the variance of diverse estimated cross-correlation angular power spectra between mock LBGs and mock CMB lensing maps, where the LBG maps were either contaminated or corrected and (ii) the un-contaminated variance.}
\label{fig:variance_mock}
\end{figure*}

\section{Regressis weights}
\label{app:regressis_weights}
\begin{figure*}
    \centering
        \includegraphics[width=0.8\linewidth]{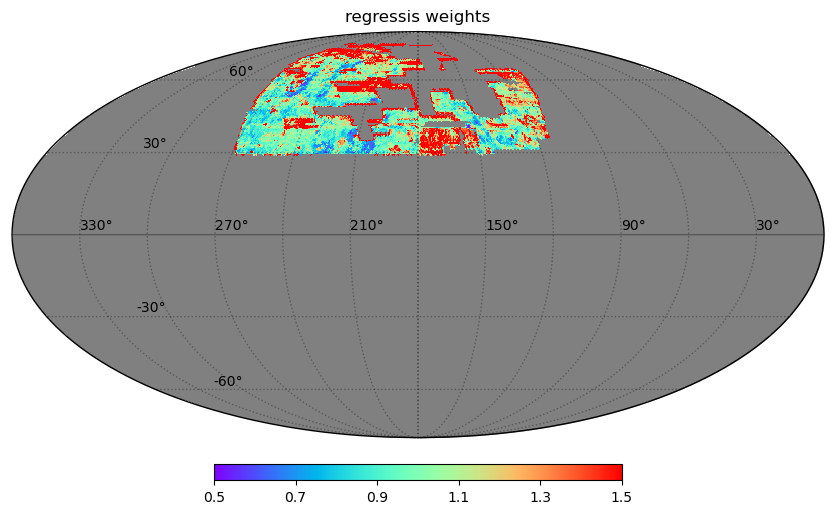}
        \caption{Random Forest \texttt{regressis} systematic weights.}
    \label{fig:wRF_map_diagnostic}
\end{figure*}

\begin{figure*}
    \centering
\includegraphics[width=0.9\linewidth]{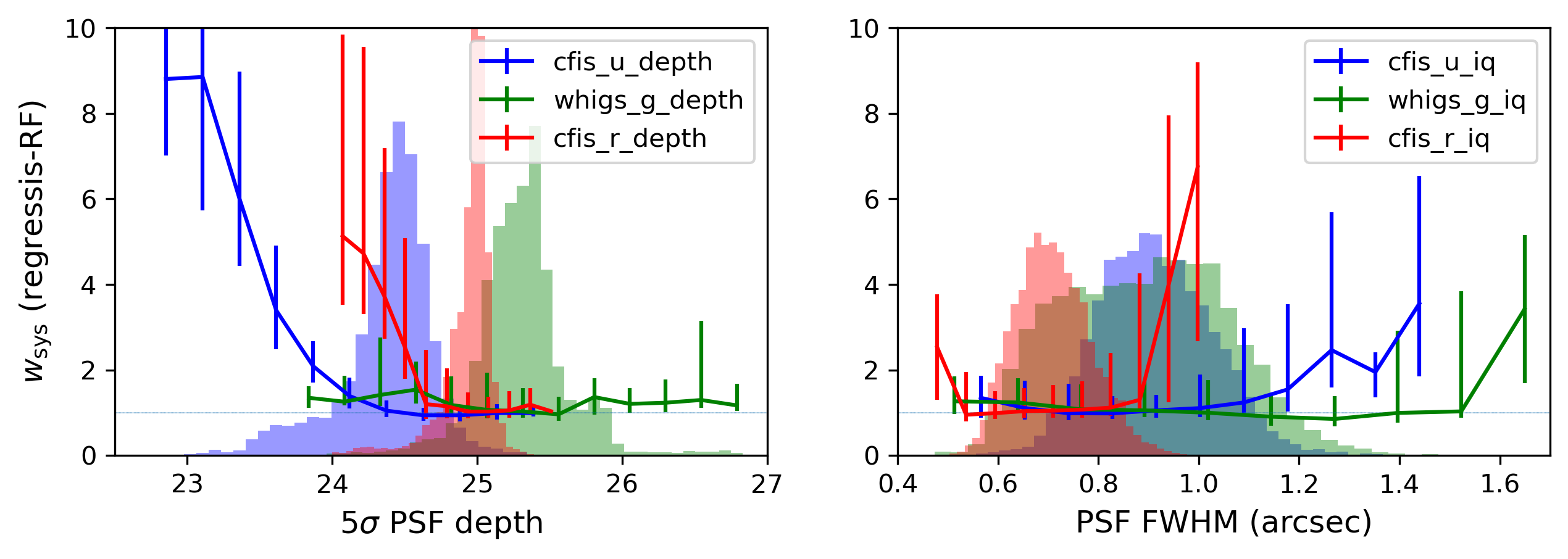}
        \caption{\texttt{regressis} systematic weights versus UNIONS imaging features (left: $5\sigma$ PSF depth, right= PSF size).}
    \label{fig:wRF_diagnostic}
    \centering
\end{figure*}

\begin{figure*}
    \centering
\includegraphics[width=1\linewidth]{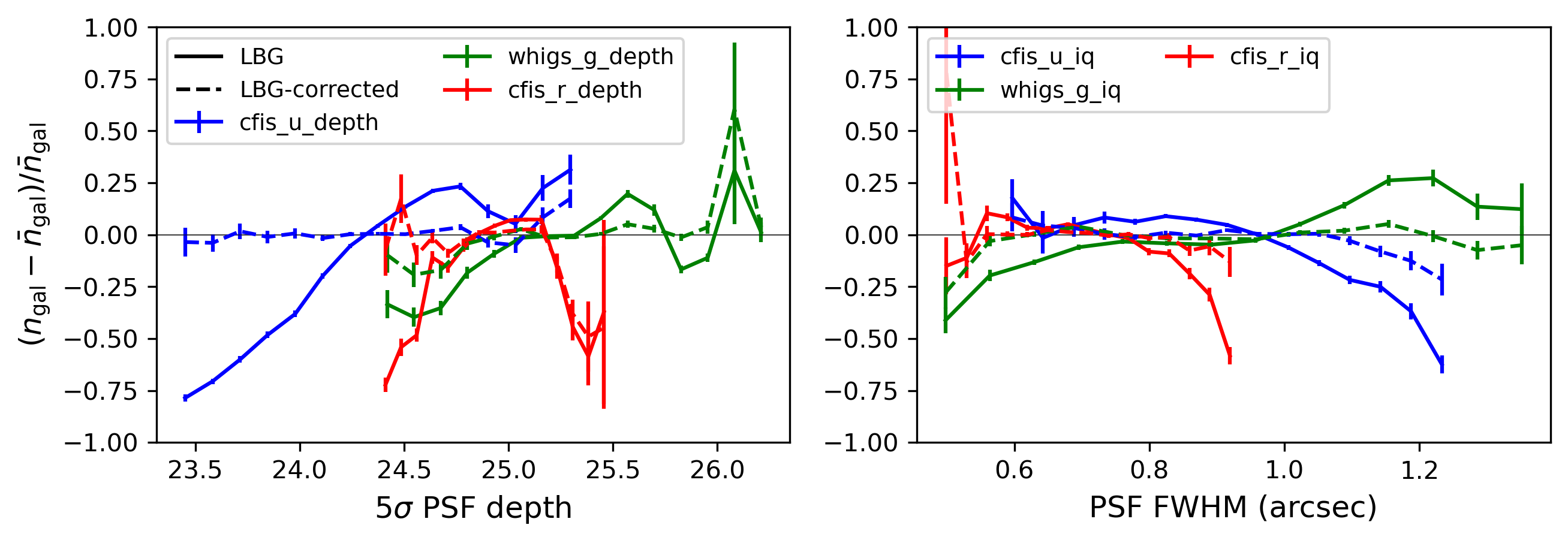}
\includegraphics[width=1\linewidth]{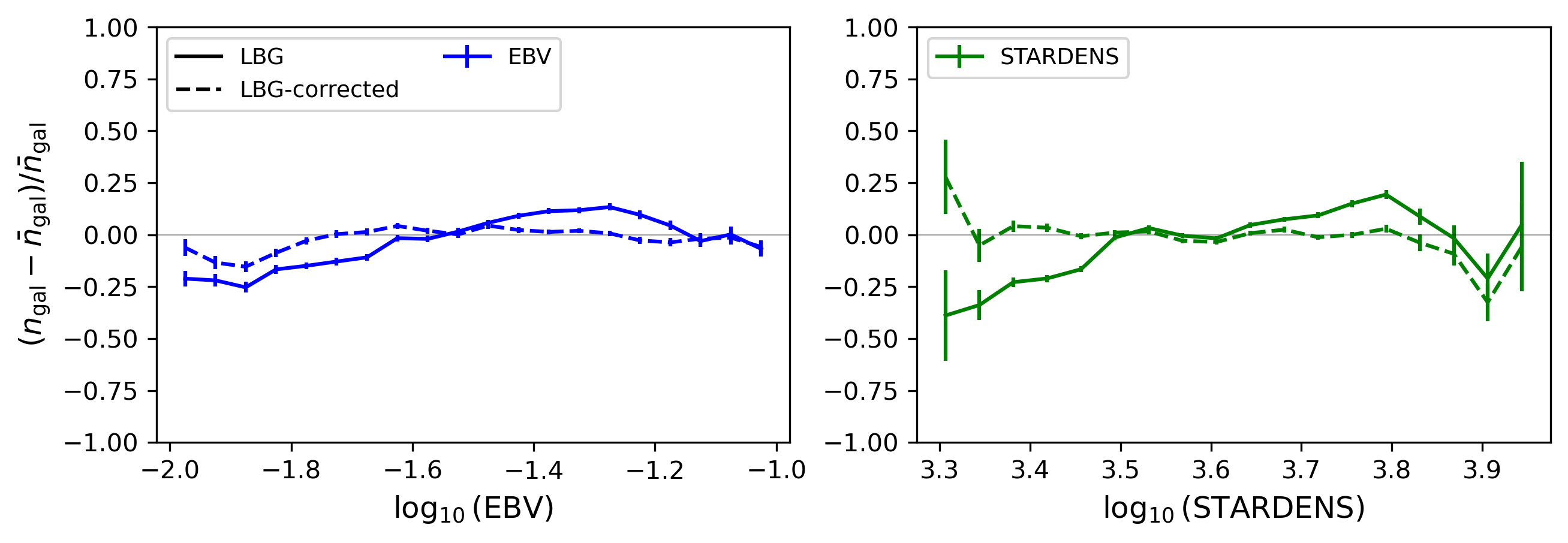}

        \caption{Same as Figure \ref{fig:overdensity_VS_template_namaster}, but with \texttt{regressis} RF correction method.}
    \label{fig:wRF_diagnostic_per_template}
    \centering
\end{figure*}

Figure \ref{fig:wRF_map_diagnostic} displays the RF \texttt{regressis} weights obtained by considering the 6 UNIONS templates and 2 external templates, fitted on the UNIONS LBG data. Figure \ref{fig:wRF_diagnostic} shows the same weights, binned in intervals of imaging templates. As for the \texttt{NaMaster} plot in \ref{fig:overdensity_VS_template_namaster}, Figure \ref{fig:wRF_diagnostic_per_template} shows the LBG overdensity before and after \texttt{NaMaster} correction.

\begin{figure*}
    \centering
\includegraphics[width=1\linewidth]{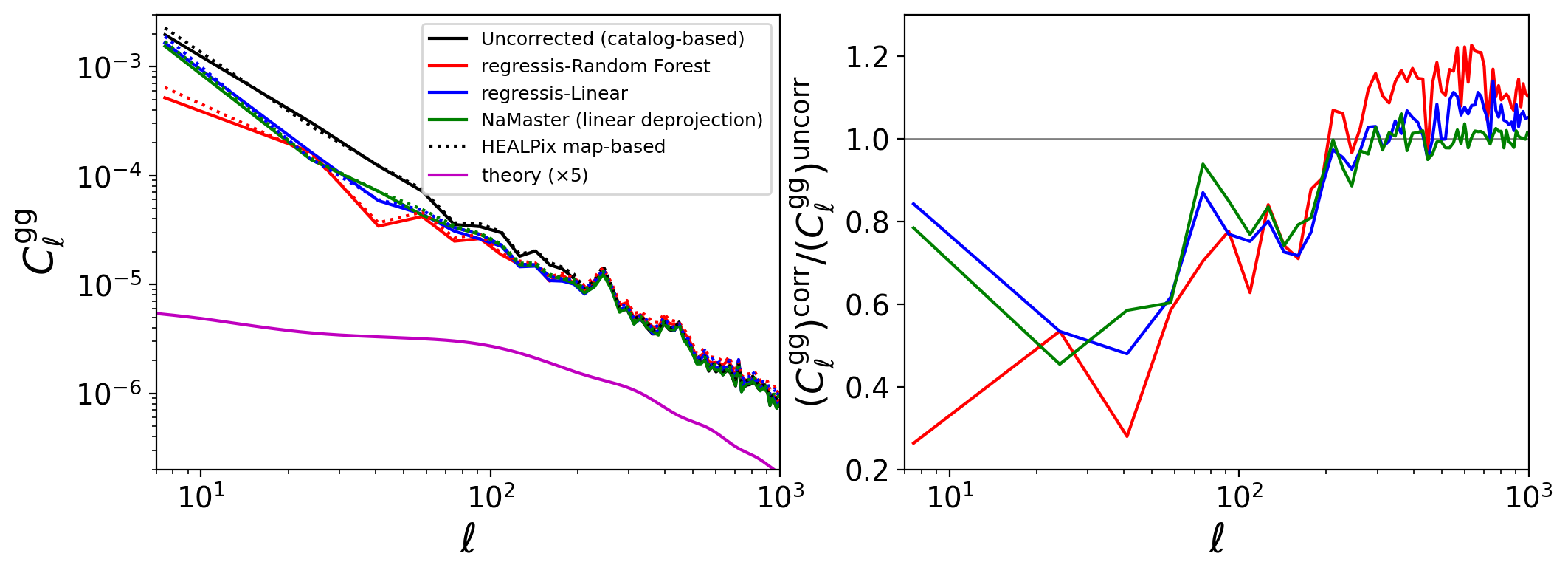}
\caption{Same as Figure \ref{fig:Angpow_Cell_gg_lbgudropout}, with additional corrected spectra with \texttt{regressis} (either Linear or Random Forest).}
    \label{fig:regressis_Cell}
    \centering
\end{figure*}

\section{Jackknife covariance: validation with mocks}
\label{app:jackknife_validation_with_mock}
\begin{figure*}
    \centering
    \includegraphics[width=0.5\linewidth]{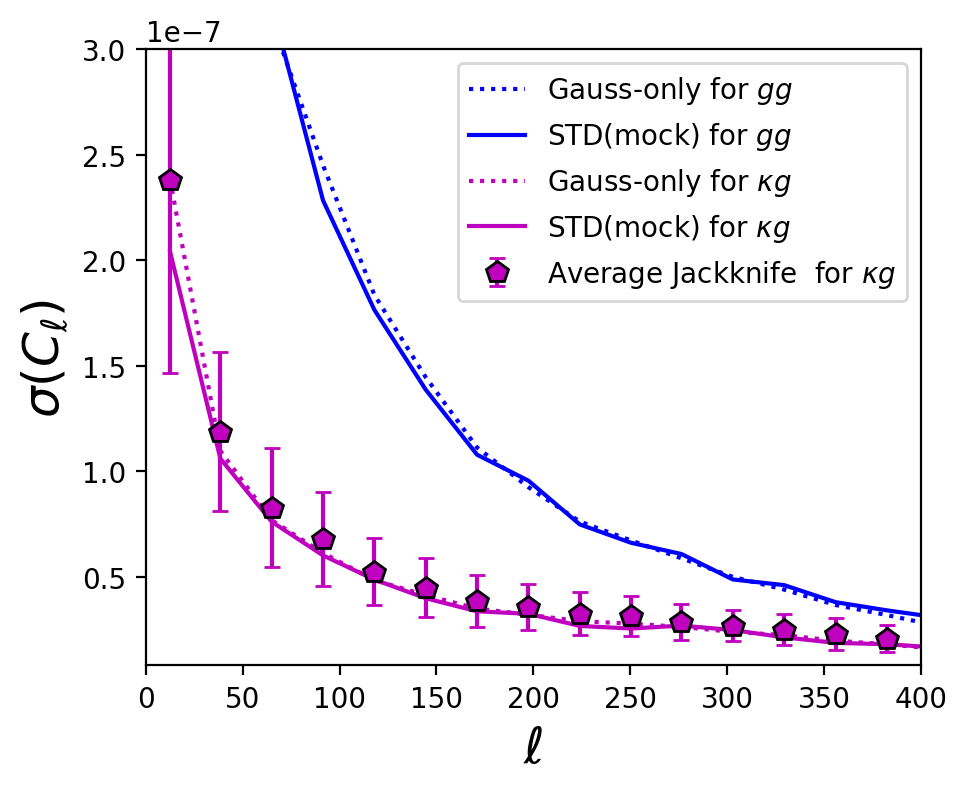}
    \caption{Variance of the auto and cross-spectrum (theory, standard deviation over 100 simulations, Jackknife).}
    \label{fig:variance}
\end{figure*}

Figure \ref{fig:variance} shows the diagonal component of the angular power spectrum (either auto or cross). It is first given as a prediction in Eq. \eqref{eq:covariance_Gauss_pred}. It is validated by computing the variance of the estimator over 100 simulations, with the Jackknife method. 
\end{appendix}
\end{document}